\documentclass{jfm}
\usepackage{graphicx}
\usepackage{stmaryrd}
\usepackage{epstopdf, epsfig}
\usepackage{placeins}
\usepackage{hyperref}
\usepackage{makecell}
\usepackage{xcolor}

\usepackage{tikz}

\shorttitle{Non-linear dynamics in the wake of a floating offshore wind turbine}
\shortauthor{T. Messmer, M. Hölling and J. Peinke}

\title{Enhanced recovery and non-linear dynamics in the wake of a model floating offshore wind turbine submitted to side-to-side and fore-aft motion}

\author{Thomas Messmer\aff{1}
  \corresp{\email{thomas.messmer@uni-oldenburg.de}},
  Michael Hölling\aff{1},
  \and Joachim Peinke\aff{1}}

\affiliation{\aff{1}ForWind - Center for Wind Energy Research, University of Oldenburg, Institute of Physics, 26129 Oldenburg, Germany}

\begin{document}

\maketitle

\begin{abstract}
An experimental study in a wind tunnel is presented to explore the wake of a floating wind turbine subjected to harmonic side-to-side and fore-aft motions under laminar inflow conditions. The wake recovery is analysed as a function of the frequency of motion, measured by the Strouhal number,  $St$. Our findings indicate that both directions of motion accelerate the transition to the far-wake compared to the fixed turbine. The experimental outcomes confirm the CFD results of \cite{li2022onset} showing that sideways motions lead to faster wake recovery, especially for $St \in [0.3, 0.6]$. Additionally, we find that fore-aft motions also lead to better recovery for $St$ $\in [0.3, 0.9]$.
Moreover, we see that the recovery is closely linked to non-linear spatio-temporal dynamics. Special non-linear dynamical effects are found in the shear layer region of the wake ($y \approx R$). For both directions of motion and $St \in [0.25, 0.55]$, the noisy wake dynamics synchronise to the frequency of the motion. For fore-aft motions and $St \in [0.55, 0.9]$, the wake shows more complex quasi-periodic phenomena. In fact, the driven frequency and a self-generated meandering mode emerge and interact non-linearly, as proved by occurring mixing components. Sideways motions result in large spatial structures of meandering. Fore-aft motions induce pulsing of the wake. Both modes are linked to the synchronisation effect. The quasi-periodic phenomena of fore-aft motions are connected to meandering mode. Overall, the spatio-temporal phenomena lead to faster recovery likely due to an increase of momentum transport towards wake centre.
\end{abstract}

\begin{keywords}
Authors should not enter keywords on the manuscript, as these must be chosen by the author during the online submission process and will then be added during the typesetting process (see http://journals.cambridge.org/data/\linebreak[3]relatedlink/jfm-\linebreak[3]keywords.pdf for the full list)
\end{keywords}

\section{Introduction}
\label{sec:intro}

The worldwide installed capacity of offshore wind is significantly growing since 2010. Most of the current offshore wind turbines are mounted on fixed structures and are clustered in wind farms installed in water depths lower than 60 meters \citep{diaz2020review}. For water depth larger than 60 meters, floating wind is the preferred solution \citep{hannon2019offshore}. Compared to a fixed turbine, a floating offshore wind turbine (FOWT) is free to move in its 6-degree-of-freedoms (DoFs). The action of wind, waves and current is responsible of complex platform motions \citep{jonkman2011dynamics}. These motions depend among others on platform types, environmental conditions, moorings and water depth \citep{butterfield2007engineering, goupee2014experimental}. The effect of these motions on the aerodynamic of the rotor and the development of the wake is an active research topic becoming highly relevant for wind farms layout. 
\\
\\
Typical distances between offshore turbines within a farm in the main direction of the wind are in the range of 6D to 12D  \citep{hou2019review}. This implies that downstream turbines operate in the wake of upstream turbines. The wake of an offshore turbine is a high-turbulent and low energy flow region compared to the undisturbed flow. There is thus a lot of interest to understand the evolution of the wake of a wind turbine and especially the features of the far-wake (typically for $x/D \geq 6$). Over the past decades, several numerical studies, theoretical models, field measurements and wind tunnel experiments have been carried out, leading to significant progress in the rich subject of fixed wind turbine wakes \citep{ainslie1988calculating, vermeer2003wind, chamorro2009wind, porte2020wind, neunaber2020distinct}. 
\\
\\
\indent  (i) \textit{Wake regions of a fixed wind turbine}. As the study of the wake of a FOWT prerequisites a good knowledge of the wake of a fixed turbine, we start a discussion of wake features of a fixed turbine.  Recent work by \citet{neunaber2020distinct} depicts in great details the different regions in the wake generated by a wind turbine. The \textbf{near-wake} is located in the vicinity of the turbine, characterised by the presence of hub-, root- and tip-vortices \citep{vermeer2003wind}. These vortices are advected downstream by the mean flow, until instabilities occur, which lead to the destruction of the vortex system. The complex mechanisms that lead to the destruction of the coherent structures are extensively described in \citep{widnall1972stability, felli2011mechanisms, lignarolo2015tip}. \cite{lignarolo2015tip} confirmed the hypothesis of \cite{medici2005experimental} regarding the fact that the tip-vortices shield the wake from the outer flow and prevent the exchange of momentum that provides re-energisation to the wake. The process of recovery thus starts when the tip vortices become unstable, which marks the beginning of the \textbf{transition region}. At that point, the shear layer which separates the wake to the undisturbed flow grows radially. Momentum is transported from the outer region to the wake and turbulence builds up in the wake. When the shear layers merge at the center, the amount of turbulence in the wake is at its maximum and then gradually decrease as moving downstream (similar to the behaviour of the wake of a fractal grid as shown by\cite{hurst2007scalings}). At this specific point, the wind speed along the hub center line begins to recover. This area is the \textbf{decay region}. Finally, when turbulence in the wake has reached an homogeneous-isotropic state \citep{pope2000turbulent}, the \textbf{far-wake} region kicks-off. The size of each region, the intensity of the recovery and the turbulence of the wake depends greatly on the type of inflow and the operating conditions of the turbine \citep{wu2012atmospheric, iungo2013field, neunaber2017comparison}. It is clear that the emergence of the far-wake are directly linked to the phenomena that happened in the regions closer to the rotor, however in the center of the far-wake the detailed features of its turbulence seem to be universal. The breakdown of tip-vortices generate small-scale turbulence that transports momentum from the surroundings to the wake, thus being an important mechanism for the recovery of the wake \citep{lignarolo2015tip}.
\\
\\
\indent  (ii) \textit{Wake meandering of a fixed wind turbine}. An important property of the wake of a wind turbine is its tendency to oscillate mainly side-to-side, so-called wake meandering \citep{medici2006measurements, larsen2008wake, howard2015statistics}. This phenomenon originates from two different causes: on the one hand, the turbulent flow of the atmosphere contains large eddies whose scale is larger than the wake width. These large eddies pass through the rotor and are responsible for the low frequency and large amplitude oscillation of the wake \citep{larsen2008wake, espana2011spatial, heisel2018spectral}. On the other hand, wake instabilities can cause meandering in the far-wake, where the wake tends to oscillate at a frequency, $f_m$ expressed in terms of Strouhal number, $St_m = f_m D/U_{\infty} \in [0.1, 0.5]$. Observations from various studies, such as \cite{okulov2014regular, foti2018similarity, heisel2018spectral, gupta2019low}, have revealed a broad peak in the spectra at $f_m$, which indicates that wake oscillations differ from classical vortex shedding, which typically exhibits a sharp peak at the shedding frequency. After \cite{gupta2019low} wake meandering occurs in the far-wake due to the ``amplification of upstream disturbances by shear flow instabilities". Perturbations around the natural frequency of wake meandering, $f_m$ can cause an early onset of meandering \citep{mao2018far, gupta2019low}. 
\\
\\
\indent (iii) \textit{Wake of a floating wind turbine}. The latest topic of floating wind turbines raises fundamental questions about the impact of movements on the development of the wake. Another aspect not further discuss here is the impact on rotor aerodynamic. Firstly, the motions of a floating turbine are responsible for unsteady aerodynamic phenomena. Unsteadinesses are observed for high frequency of motions, characterised by the dimensionless Strouhal number,  $St = f_p D / U_{\infty}$ (where $f_p$ is the frequency of motion, $D$ is the rotor diameter, and $U_{\infty}$ is the inflow speed), typically for $St > 0.5$ \citep{sebastian2013characterization, farrugia2014investigating, sant2015measurements, bayati2017formulation, fontanella2021unaflow}.
Secondly, rotor motions impact wake dynamics, which was shown through various wind tunnel studies with model turbines and porous disks \citep{rockel2014experimental, rockel2017dynamic, bayati2017wind, fu2019wake, schliffke2020wind, kopperstad2020aerodynamic, fontanella2022wind, meng2022wind, belvasi2022far}. \cite{bayati2017wind} discussed the relevance of the so-called  ``wake reduced velocity", which is equivalent to a Strouhal number, for the  characterisation of unsteady aerodynamic conditions. \cite{fu2019wake} measured with PIV the wake of a pitching and rolling model turbine with rather high amplitudes (up to $20^{\circ}$) an low frequencies ($St < 0.03$). They observed a stronger recovery for the floating turbine and enhanced Turbulent Kinetic Energy (TKE), which was also observed by \cite{rockel2017dynamic}. A more systematic study was carried out by \cite{schliffke2020wind} in which they measured at 4.6D downstream the wake of a surging porous disc with $St$ $\in [0, 0.24]$ under realistic turbulent conditions. No impact of motions on the recovery of the wake was measured, however a signature of the motions in the spectra of the wake was observed for all frequencies. \cite{kopperstad2020aerodynamic} investigated in a wind tunnel and with CFD the wake of a spar and barge floating turbine. They showed that in a laminar wind, the wake of a floating turbine recovers faster compared to a fixed machine, for which the recovery depends on the wave excitation. Numerical simulations using free-wake vortex code and CFD were carried out to study the different wake regions of a floating turbine \citep{farrugia2016study, lee2019effects, chen2022modelling, kleine2022stability, chen2022modelling, li2022onset, ramos2022investigation}. \cite{farrugia2016study} outlined the possibility for surge motion ``to induce the onset of complex interactions between adjacent tip vortices". These instabilities induced by surge motion on the complex helical vortex system could explain the faster recovery of the wake of a surging turbine for a range of $St$  $\in [0.4, 1.7]$, studied by \cite{ramos2022investigation}. Based on stability theory of vortices, \cite{kleine2022stability} showed that the motion of a floating turbine ``excites vortex instabilities modes", which were predicted by \cite{farrugia2016study}. They demonstrated that motions with a frequency of $0.5$ and $1.5$ the rotor 1P frequency induce the strongest disturbances. \cite{chen2022modelling} carried out CFD simulations of the wake of a surging turbine with amplitude $A_p \in [0.03D, 0.1D]$, $St \in [0.41, 1.64]$. They showed the positive impact of motions on the recovery of the wake of a surging turbine, where up to 10 \% more recovery compared to the fixed turbine was found. Last but not least, \cite{li2022onset} used CFD to study the onset of wake meandering arising from the sideways motion of a floating turbine. They demonstrated that sway motions for $St \in [0.2, 0.6]$ helps the wake to recover up to 20 \% more compared to fixed case. They showed that side-to-side turbine motions lead to large wake meandering, even with small amplitude of motions ($A_p/D \approx 0.01$). This important result is particularly pronounced for low turbulent conditions ($TI < 4 \%$). 
\\
\\
The above mentioned discussion shows that the current understanding of the wake of floating wind turbines is based on many specific investigations. In \cite{phdseminar2021} we reported on p.4-5 on wind tunnel experiments that we conducted in 2021 on cases similar to \cite{li2022onset}. In this paper, we extended our experiments to side-to-side and fore-aft harmonic motions to obtain a more global understanding of the wake development for different types of movements of a FOWT. A special focus is given to the influence of the frequency of movements. In order to examine solely the impact of the motion on the wake and to exclude any impact of inflow turbulence, the turbine was operated in laminar wind. In this work, we addressed the following questions:
\\
\begin{itemize}
\item[(i)] Which non-dimensional numbers and principle directions of motion drive the wake dynamics of a floating wind turbine?
\item[(ii)] How do side-to-side and fore-and-aft movements affect the development of the far wake
of a floating turbine? Is there a range of motion frequencies that allows the wake of a moving turbine to recover faster? What are the underlying mechanisms?
\end{itemize}
\  \\
The \S \ref{sec:exp} details the experimental set-up, the cases investigated and the methodology for the analysis. The \S \ref{sec:results} presents the averaged results, i.e the profiles of wake deficit and turbulence intensity. Here we show the dependency to the DoF on the wake. We describe the effects of side-to-side motions with $St \in [0, 0.58]$ and fore-aft motions with $St \in [0, 0.97]$ on the wake recovery. The \S \ref{sec:discussion} examines the dynamics of the wake and discusses the results in terms of non-linear dynamics.

\section{Experiments}
\label{sec:exp}

\subsection{Set-up}

\begin{figure}
  \centerline{\includegraphics[width=13cm]{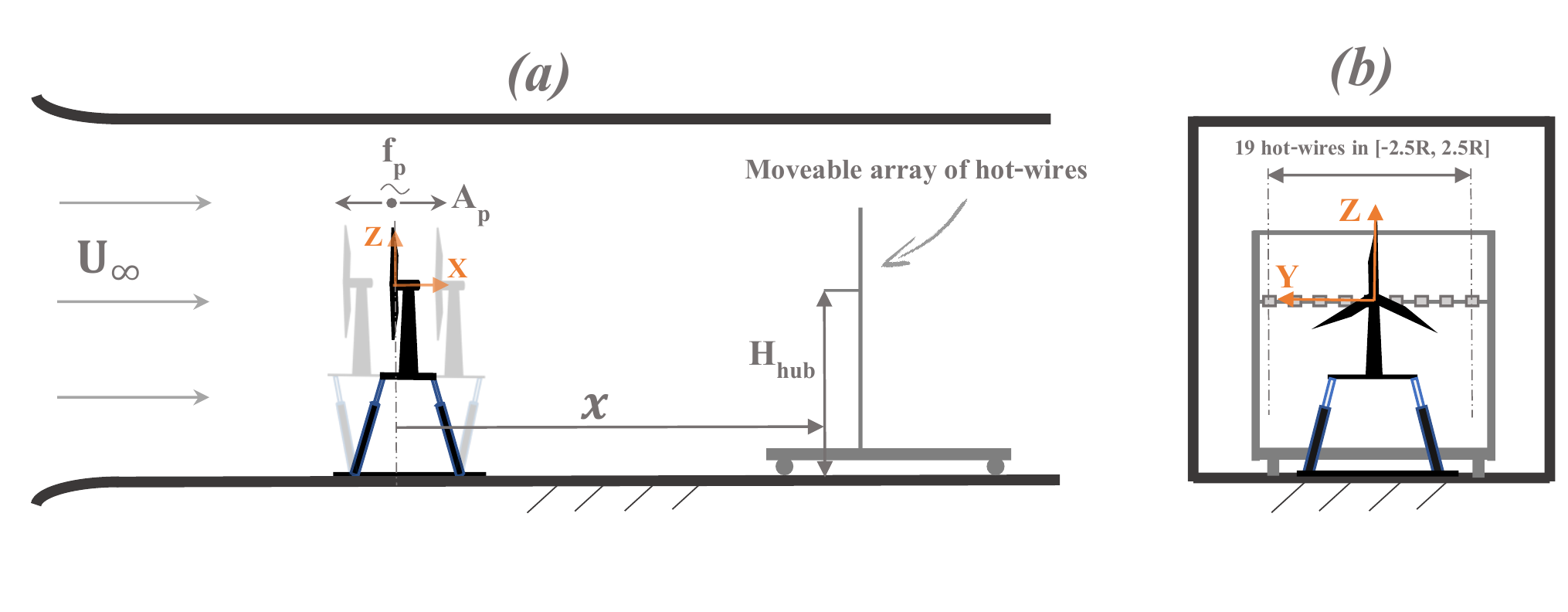}}
  \caption{Scheme of the experimental set-up, the MoWiTO 0.6 mounted on a 6-DoF platform installed in the wind tunnel
  (\textit{a}) side view and (\textit{b}) front view. \textit{not to scale}}
\label{fig:set_up}
\end{figure}

The experimental set-up used to perform the investigations is shown in figure \ref{fig:set_up}. The experiments were done in the $3 \times 3~m^2 \times 30~m$ test section of the large wind tunnel of the University of Oldenburg \citep{kroger2018generation}. The set-up consists of the Model Wind Turbine Oldenburg 0.6 (MoWiTO 0.6) with a diameter of $\sim$0.6 m mounted on a motorised platform. 

A 6-DoF motorised platform was designed specifically for the MoWiTO 0.6 in order to investigate the aerodynamic of FOWTs \citep{messmer2022six}. With this system, the model turbine can be moved following pre-defined motion signals in the 6 degree-of-freedoms, namely the three translations: surge, sway and heave and the three rotations: roll, pitch and yaw. The model turbine was equipped with a strain gage to measure the thrust, $T$, of the ensemble \{tower + rotor\}. For any case investigated: power produced, rotational speed and thrust of the turbine was recorded. The movements of the platform were recorded to verify the adequacy of the motions. The wind speed, $U$, in the wake of the model turbine was measured with an array of 19 1-dimensional hot-wires of $\sim$1mm length, resolving turbulent scale of this size.
\\
The probes were operated with multichannel 54N80-CTA (Constant Temperature Anemometer) modules from \textit{Dantec Dynamics}. Data were acquired with a sampling rate of 6 kHz, gathering $\sim 10^6$ points for each measurement. The time of measurements, $T_{meas}$ was sufficient to reach statistical convergence of any post-processed data shown and discussed in this paper. The hot-wire probes were calibrated twice daily and temperature, humidity and air pressure were recorded throughout the day to track any drift effects, which came out to be negligible. The inflow wind speed was measured with a Prandtl-tube about two meters in front of the model turbine.
\\
The array was mounted on a moving motorised cart which allowed to measure the wind speed, at the downstream position, $x$, from 6D to 10D (cf figure \ref{fig:set_up} (a)). We selected this range of distances based on the typical layout of a wind farm as mentioned in \S \ref{sec:intro}.
The hot-wires were placed on a horizontal line at hub-height (around 1 meter above floor level) and span lateral positions, $y$ from $-2.5R$ to $2.5R$ ($R = D/2$)  with respect to the hub centre (cf figure \ref{fig:set_up} (b)).

\subsection{Model Wind Turbine Oldenburg 0.6}

The MoWiTO 0.6 used for the experiments has a rotor diameter, $D$ of 0.58 m. The turbine is developed at the University of Oldenburg  \citep{schottler2016design, juchter2022reduction}. The properties of the turbine are depicted in appendix \ref{appA} in table \ref{tab:mowito_06_table}. 
The model turbine was used for several measurement campaigns with a focus on wake flow, further details of the experimental procedure and wake measurements can be found in \cite{hulsman2020turbine, neunaber2020distinct, neunaber2022leading}.

\begin{figure}
  \centerline{\includegraphics[width=13cm]{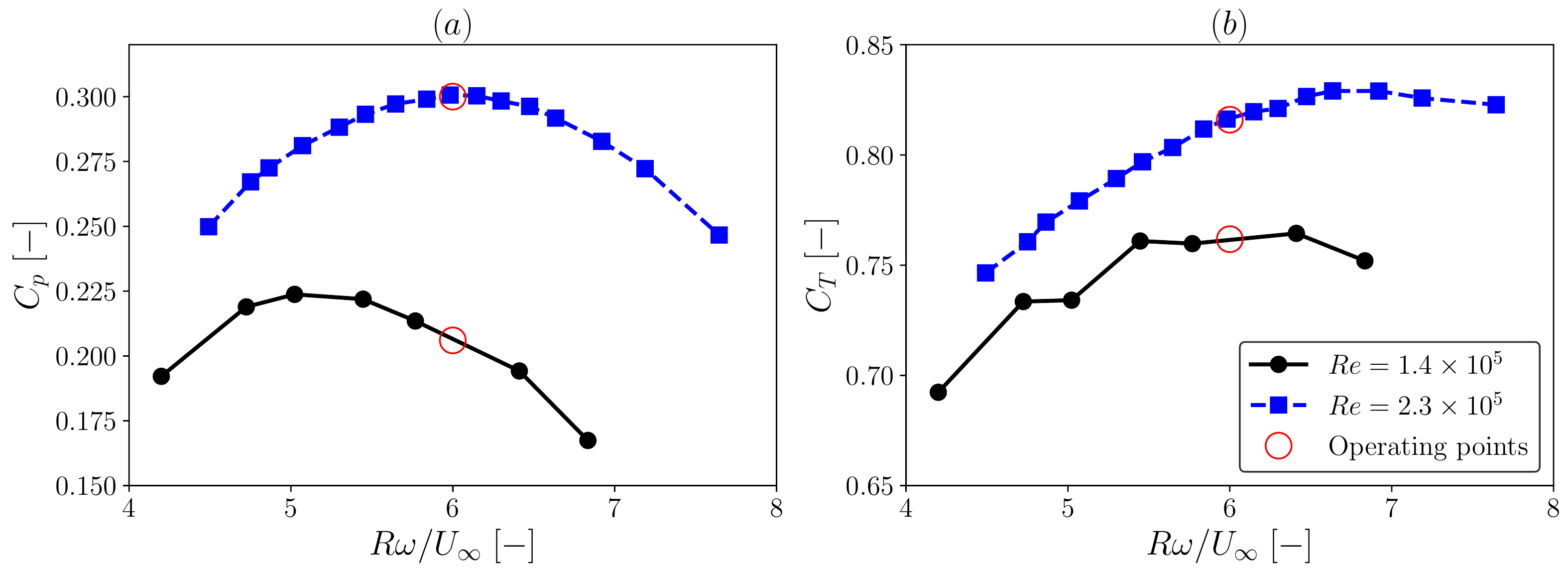}}
  \caption{$C_P$ (a) and $C_T$ (b) versus tip speed ratio of the MoWiTO 0.6 for two rotor-based Reynolds number (for $U_{\infty} = 3, ~5$ m/s) at fixed blade pitch angle}
\label{fig:cp_ct_lambda}
\end{figure}

The turbine used for the measurements was characterised by power and thrust coefficients ($C_P$, $C_T$) vs Tip Speed Ratio ($TSR$). Figure \ref{fig:cp_ct_lambda} (a) displays $C_p$ vs. $TSR$ and  figure \ref{fig:cp_ct_lambda} (b) $C_T$ vs. $TSR$ for $U_{\infty} = 3, ~5$ m/s resulting in a rotor-based Reynolds number, $\Rey$ from $1.4 \times 10^5$ to $2.3 \times 10^5$. The power coefficient is $\Rey$ dependent. For $TSR \approx 6$, $C_p$ increases from 0.2 at low $\Rey$ to 0.30 at the highest $\Rey$. This is consistent with the features of the low-$\Rey$ airfoil used, SD7003, which performs better at high $\Rey$ \citep{selig1995summary}. The thrust coefficient is around $0.8$ for $TSR \approx 6$. The thrust coefficient account for rotor and tower aerodynamic loads. In a previous study by \cite{neunaber2020distinct}, the load on the tower and nacelle of the MoWiTO 0.6 was measured without the blades. The authors assessed this value to be 17 \% of the total thrust coefficient. The operating points of the wake experiments are marked by red circles in figure \ref{fig:cp_ct_lambda}. All measurements were carried out at $TSR \approx 6$ for which $C_T \approx 0.76$ at 3 m/s and $C_T \approx 0.81$ at 5 m/s.

\subsection{Non-dimensional parameters that drive the wake behaviour}
\label{sub-sec:dimensionless}

As for our desired experiment investigations many quantities are involved, we briefly discuss the selection of necessary parameters for a complete characterisation. Therefore we use the basis of the $\pi$-theorem. For a given DoF, the system investigated involves a total of nine independent parameters, namely platform motion frequency and amplitude, inflow wind speed, rotor diameter, air viscosity, air density, turbine thrust and power and rotational speed. These are noted: $f_p$, $A_p$, $U_{\infty}$, $D$, $\mu$, $\rho$, $T$, $P$, $\omega$. The $\pi$-Theorem suggests that, a problem characterised by m dimensional variables can always be reduced to a set of m-n dimensionless parameters ($\pi$-groups), with n
the fundamental units of measure (dimensions) as depicted by \cite{buckingham1914physically}. Therefore this problem with nine variables and three dimensions can be reduced to six dimensionless parameters. These are all defined in table \ref{tab:non-dim_param}, namely the thrust coefficient, $C_T$, power coefficient, $C_P$, tip speed ratio, $TSR$, reduced amplitude, $A^*$, Strouhal number, $St$, and Reynolds number, $\Rey$. Thus in a laminar flow, the properties of the wake of a FOWT, $wake_{FOWT}$, are determined by these six dimensionless numbers: $wake_{FOWT} = f(C_T,~C_P,~TSR, ~A^*,~St,~Re)$.

\begin{table}
  \begin{center}
\def~{\hphantom{0}}
  \begin{tabular}{cccccc}
  		\makecell{ Thrust \\ coef. } &  \makecell{ Power \\ coef. } &  \makecell{ Tip Speed \\ Ratio }  &  \makecell{ Reduced \\ amplitude }  & \makecell{ Strouhal \\ number } &  \makecell{ Reynolds \\ number} \\
  		\\
$C_T$ &  $C_p$ &  $TSR$ &  $A^*$  & $St$ &  $\Rey$\\
\\
$\frac{T}{1/2 \rho \pi R^2 U^2_{\infty}}$ &  $\frac{P}{1/2 \rho \pi R^2 U^3_{\infty}}$ &  $\frac{R \omega}{U_{\infty}}$ &  $\frac{A_p}{D}$  & $\frac{f_p D}{U_{\infty}}$ &  $\frac{\rho U_{\infty} D}{\eta}$\\
  \end{tabular}
  \caption{Dimensionless parameters that drive the wake of a FOWT}
  \label{tab:non-dim_param}
  \end{center}
\end{table}

The dependency to $C_T$ of wake recovery and expansion of a wind turbine was characterised extensively \citep{porte2020wind}. The tip speed ratio also plays an important role on the development of the wake of a FOWT \citep{farrugia2016study}. Both the amplitude and frequency of motions can impact the wake of a moving turbine, as demonstrated by \cite{chen2022modelling, li2022onset, ramos2022investigation}. As already mentioned above, motion frequency has a greater influence on wake dynamics, even at low amplitudes ($A_p \sim 0.01D$). Based on these works and the set of parameters, we concluded that it is worth to focus on the impact of different $St$ at constant amplitude of motions.

\subsection{Cases investigated}
\label{sub-sec:cases_investigated}

\begin{figure}
  \centerline{\includegraphics[width=12cm]{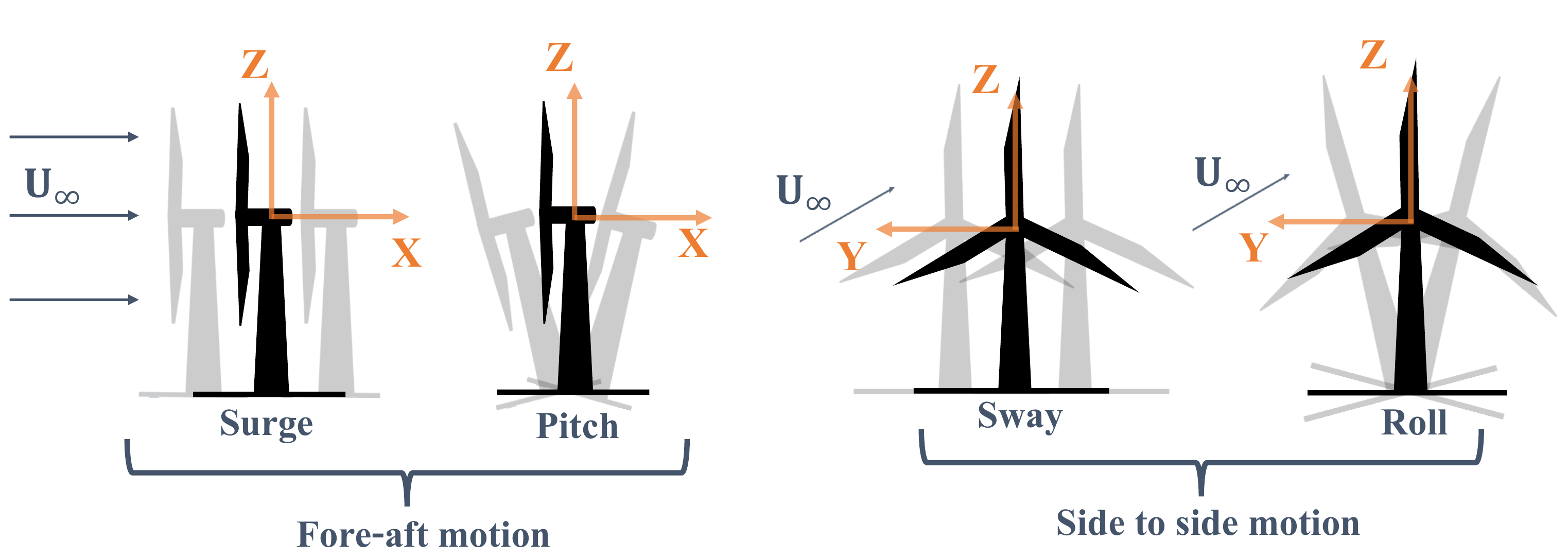}}
  \caption{Fore-aft (surge and pitch) and side-to-side motions (sway and roll) of a FOWT}
\label{fig:dof}
\end{figure}

Floating wind turbines operate in the atmospheric boundary layer, which exhibits various turbulent conditions, typically $TI \in [2, 15]$ \citep{ jacobsen2021influence, angelou2023revealing}.
The complex conditions (6-DoF motions and turbulent conditions) of a floating wind turbine were idealised for the experiments. We considered laminar wind with a low background turbulent intensity of $\sim 0.3 \%$, and imposed 1-DoF harmonic motions. Our study focused on investigating the four following DoFs: surge, sway, roll, and pitch for a specific range of $St$. Figure \ref{fig:dof} illustrates the four DoFs, with fore-aft and side-to-side motions. We examined these DoFs independently, without combining them. We imposed for a given DoF the following motion signal, $\xi$, on the platform: $\xi(t) = A_p \sin(2 \pi f_p t)$. Here, $A_p$ denotes the amplitude of motion (in meters or degrees), and $f_p$ represents the frequency of motion (in Hz).

In order to investigate a relevant range of $St$, we looked at the typical movements of a FOWT, which we briefly present below. The motions of a floating wind turbine are greatly influenced by the type of foundation used (TLP, semi-submersible and spar). Three distinct ranges of frequencies are usually observed in the motion spectra of a floating turbine:

\begin{itemize}
\setlength\itemsep{0.5em}
\item[(i)] \textit{Wave frequency}: a floating platform is subject to ocean waves which cause motions at frequencies around 0.1 Hz (wave period around 10 s). Although the wave-induced motions are rather low in amplitude ($A^* \sim 0.01$), they are high in frequency. For a 10 MW turbine at rated wind speed, typically $St \sim 1.5$ \citep{messmer2022six}.

\item[(ii)] \textit{Rotational natural frequency}: when a floating turbine is displaced from its equilibrium position (due to a gust or a series of waves), it goes back to equilibrium as a damped harmonic oscillator at a frequency that depends on the degree of freedom. For a spar or a semi-submersible, the pitch and roll natural frequencies are typically around 0.035 Hz \citep{robertson2014definition}. For a 10 MW turbine at rated wind speed, such motions give $St \sim 0.5$ and $A^*$ up to $\sim 0.04$.

\item[(iii)] \textit{Translation natural frequency}: similar to rotational degrees of freedom, a floating turbine experiences translation motions at natural frequencies. These can be large in amplitude ($A^*$ up to $\sim 0.1$) but at low frequency ($f_p \sim 0.01$ Hz). Typical Strouhal number is $St \sim 0.1$ \citep{leimeister2018critical}.
\end{itemize}
\  \\
This discussion demonstrates that floating turbines cover range of Strouhal numbers, $St \in [0, 1.5]$. While high-frequency movements are relatively low in amplitudes ($A^* \sim 0.01$), oscillations at lower frequencies can reach amplitudes up to 10\% of the rotor diameter. To test a wide range of $St$, we conducted experiments at frequencies ranging from $0.3$ to $5$ Hz with a constant amplitude of motion around $0.01D$. We considered two inflow wind speeds: 3 and 5 m/s and set the rotational speed of the turbine using a torque controller to maintain a $TSR$ of approximately 6.0 for all cases. We covered $St \in [0, 0.97]$. Table \ref{tab:cases_investigated} in Appendix \ref{appA} provides details on all the cases investigated.
\\
\\
To quantify the differences between the cases, we took as reference the fixed case ($St = 0$). To make this study of our different cases comparable, it is important to note that the mean power and thrust of the moving turbine were the same as those of the fixed turbine. If this would not have been the case, comparing the wake flows between cases would be less meaningful, as there would be differences in mean operating conditions. 
\\
It was also confirmed by examining the wake deficit in the near-wake, which was the same for any case with the same Reynolds number, $<C_T>$, $<C_P>$, and $<TSR>$. With this we concluded that the induction factor of the turbine was the same, which was also observed by \cite{fontanella2022wind}. In particular for the fore-aft motion, such a result is not obvious as motions cause temporal variations in turbine power and thrust. We found that they do not affect the mean turbine output values, at least for such low amplitudes. However, the impact of the motions is directly related to their effects on the dynamics of the wake (added turbulence and non-linearities).

\subsection{Post-processing of measured data}
\label{sub-sec:methodo}

In order to quantify the different wake measurements, we analysed the time series of local wind velocity by one- and two-dimensional statistics and by their instantaneous flow fields, defined hereafter. The measured time series are given in the form of $U(x,y, z=z_{hub}, t)$ with $x \in  \llbracket 6D, 8D, 10D \rrbracket$, $y \in [-2.5R, 2.5R]$ and $t \in [0, T_{meas} \approx 180$ s$]$.  $<.>$ denotes a temporal average for $t \in [0, T_{meas}]$.
\\
\\
\setlength\itemsep{1em}
\textbf{Normalised wind speed deficit (wake deficit)} is quantified by 
$\Delta U(x,y) / U_{\infty} = (U_{\infty} - <U(x,y,t)>) / U_{\infty}$. In the paper, we show profiles of wind deficit, the closer the deficit to zero the more the wake has recovered.
\\
\\
\textbf{Wind recovery ($\sim$ rotor normalised averaged wind speed)} is defined as
$\tilde{U} / U_{\infty}$, where $\tilde{U}(x) = \int_{y_0 - R}^{y_0 + R} <U(x,y,t)> dy$. With $y_0$ the estimated center of the wake. This quantity gives an order of magnitude of the amount of wind speed that has recovered at a specific downstream position around the rotor area, similar quantity were calculated by \cite{chen2022modelling, li2022onset}. In this paper, we also refer to $\tilde{U} / U_{\infty}$ as the recovery.
\\
\\
\textbf{Turbulence intensity} is given by $TI(x,y) = \sigma (U(x,y,t)) / <U(x,y,t)>$
\\
\\
\textbf{Power spectra}, $\Phi_x(x,y)$, is the power spectral density of the wind speed fluctuations, $u(x,y,t) = U(x,y,t) - <U(x,y,t)>$  computed with an algorithm that uses Welch’s method.
\\
\\
\textbf{Instantaneous wind field} is based on $U(x,y, z=z_{hub}, t)$ to visualise the instantaneous wake at hub height for a specific case and downstream position. A low-pass filter with a cut-off frequency of $10$ Hz was applied to remove small-scale fluctuations from the signals for each $y$ value within the range of $[-2.5R,2.5R]$. The signals were then resampled at a frequency of $500$ Hz and plotted on a colormap (see \S \ref{sec:discussion}).
\\
\\
\textbf{Cross-correlation between $U(x,y=-1/2R,t)$ and $U(x,y=1/2R,t)$} is given by $(U_{-1/2R} \star U_{1/2R}) (\tau) = <U(y=-1/2R,t)U(y=1/2R,t+\tau)> / \sigma_{-1/2R}\sigma_{1/2R}$

\section{Results of average values}
\label{sec:results}

The scheme after which we present our results is briefly explained next. We considered four DoFs, for which wake measurements are compared. The analysis of the wakes are done for the three downstream positions ($x \in \llbracket 6D, 8D, 10D \rrbracket$). The downstream dependency of the cross stream-wise profiles of wake velocity deficit and turbulence intensity is shown. Finally the recoveries of the wakes are analysed with respect to the $St$.
\\
We begin this section by presenting, in \S \S \ref{sub-section:equivalence_DOF}, the similarity between degrees of freedom that led us to reduce the analysis to two generic DoFs namely sway and surge. 
Then, the dependency of the wake recovery on the $St$ is shown in \S \S \ref{sub-sec:side-to-side} for sway. Finally, we present corresponding results for surge motions in \S \S \ref{sub-sec:fore-aft-motion}.

\subsection{Equivalence between different degrees of freedom}
\label{sub-section:equivalence_DOF}

In a previous study from \cite{bayati2017formulation}, the authors simplified the modelling of the aerodynamics of a FOWT by linearising rotational motions into translation. We checked whether this simplification is also applicable to sway/roll motions or respectively to surge/pitch motions. 

\begin{figure}
  \centerline{\includegraphics[width=12cm]{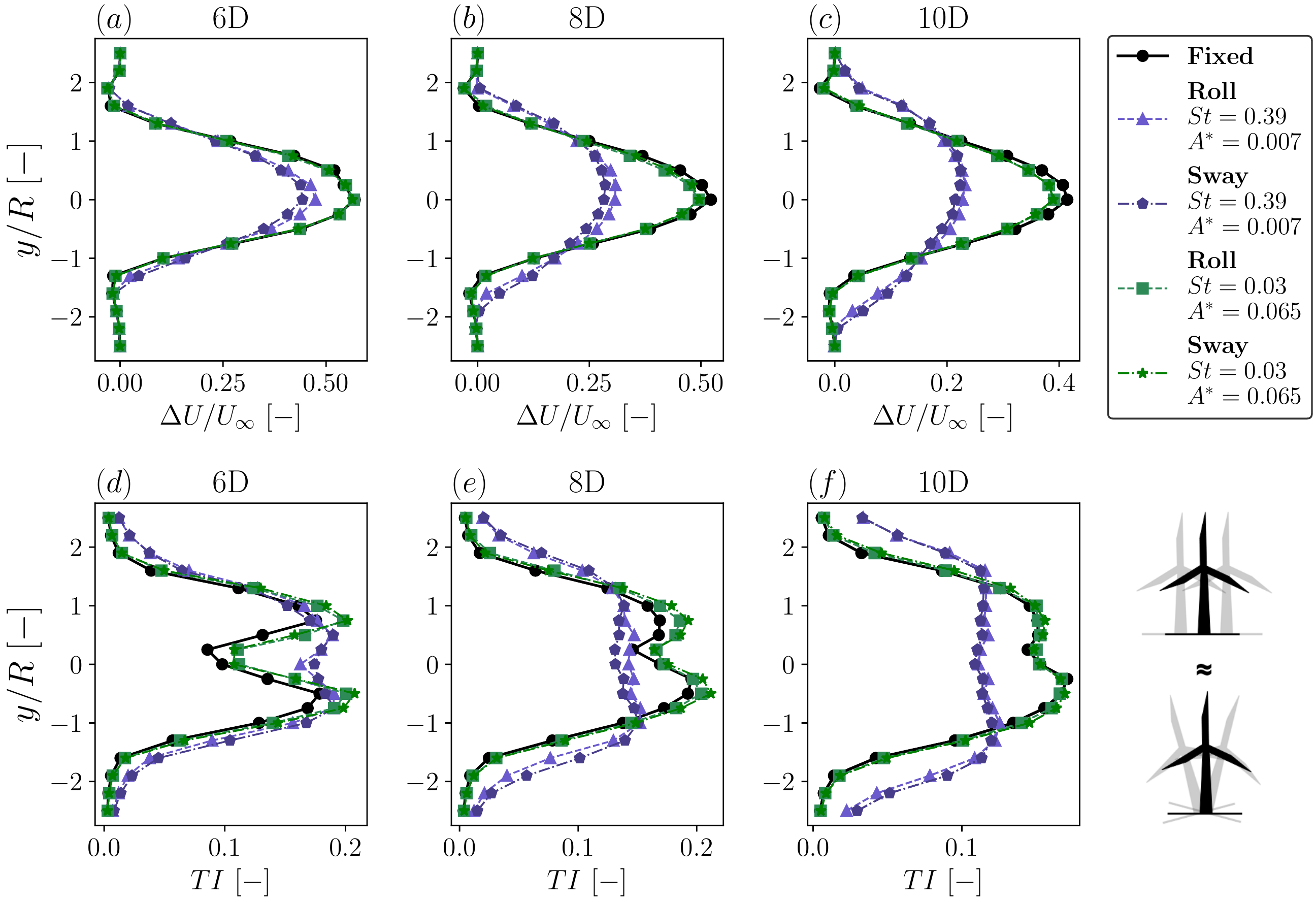}}
  \caption{Wake deficit (a-c) and $TI$ profiles (d-f) at 6D, 8D and 10D for fixed case and two \textbf{roll} and \textbf{sway} cases with same $St$ and same $A^*$, $\Rey = 2.3 \times 10^5$. Tests A.1-5 in table \ref{tab:cases_investigated}}
\label{fig:sway_roll}
\end{figure}

\begin{figure}
  \centerline{\includegraphics[width=12cm]{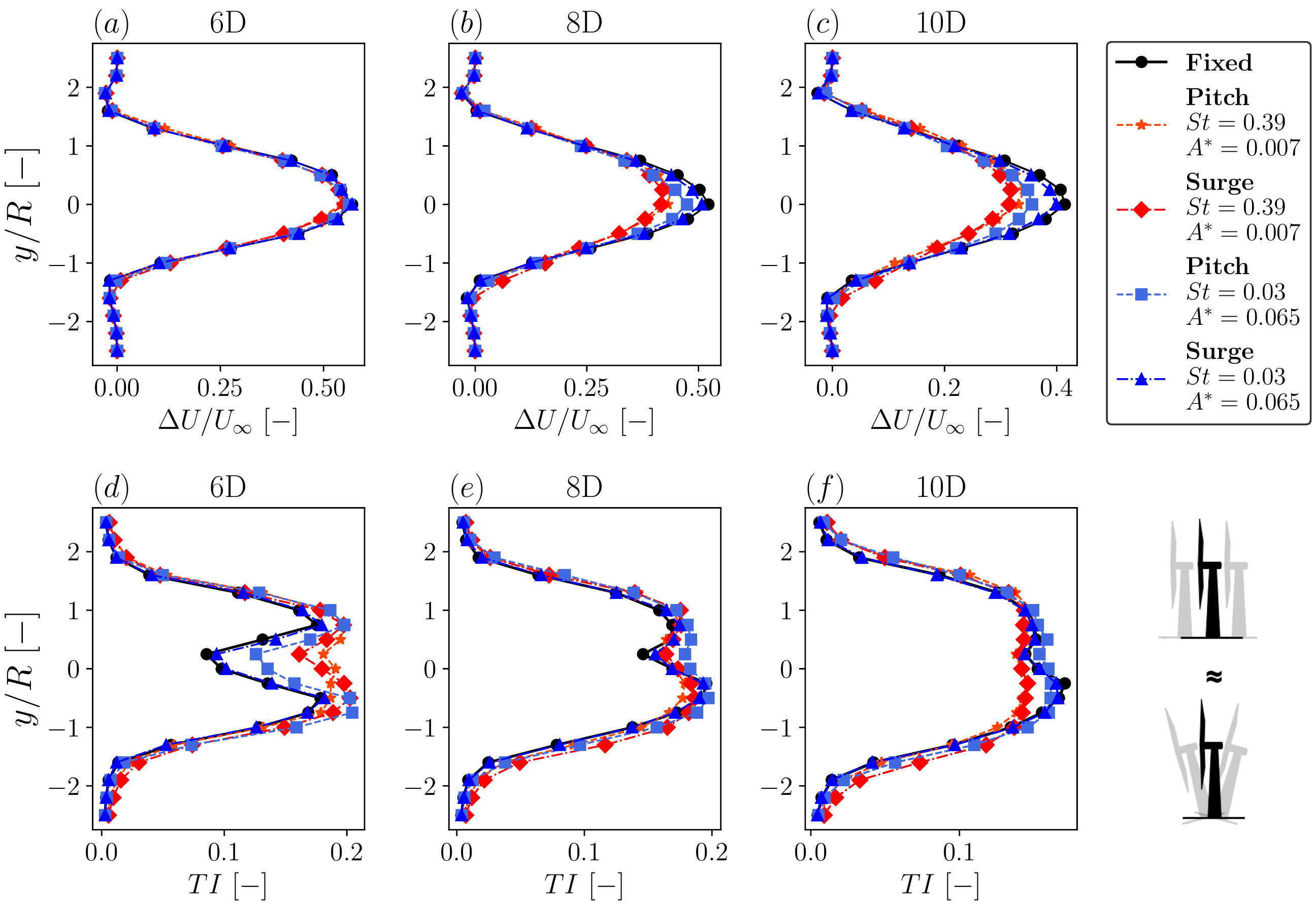}}
  \caption{Wake deficit (a-c) and $TI$ profiles (d-f) at 6D, 8D and 10D for fixed case and two \textbf{pitch} and \textbf{surge} cases with same $St$ and same $A^*$, $\Rey = 2.3 \times 10^5$. Tests B.1-5 in table \ref{tab:cases_investigated}}
\label{fig:surge_pitch}
\end{figure}

To address this issue, we measured the wake of the floating turbine with small ($A_p = 0.5^{\circ}$) and large ($A_p = 5^{\circ}$) amplitudes of platform roll around the zero value. Additionally, we performed tests with the turbine swaying at an equivalent amplitude. We carried out the same tests with pitch and surge DoFs (with no tilt angle). The results of side-to-side motions (sway and roll) are presented in figure \ref{fig:sway_roll}, while the fore-aft motions are shown in figure \ref{fig:surge_pitch}. The wind speed deficit and turbulence intensity profiles are plotted for 6D, 8D, and 10D in figures (a-c) and (d-e), respectively.
\\
For the sway and roll cases, figure \ref{fig:sway_roll} demonstrates that the match of $\Delta U / U_{\infty}$ and $TI$ is nearly perfect both low amplitudes ($A^* = 0.007$) and high amplitudes ($A^* = 0.065$). This suggests that the Bayati approximation can be applied also for the wake for these types of motions.  

Figure \ref{fig:surge_pitch} demonstrates that for low-amplitude cases ($A^* = 0.007$), the wind speed deficit and turbulence intensity profiles of the wake are very similar for pitch and surge. However, for larger amplitudes ($A^* = 0.065$), discrepancies arise. The wake of a surging turbine at low frequency ($St < 0.1$) has little impact on the wake, even at high amplitude (see figure \ref{fig:surge_pitch}, the blue curves marked with triangles follow closely the fixed case marked by black line with dotes). This is consistent with the results of \cite{meng2022wind}. In contrast, the equivalent case with pitch motion (see the blue curve with squares) clearly deviates from the fixed case. After \cite{rockel2017dynamic, fu2019wake}, the deviations observed are due to vertical wake movements induced by the rotor tilting. The wake of a pitching turbine is similar to that of a surging turbine only for low amplitudes ($A_p < 2^{\circ}$) as seen in figure \ref{fig:surge_pitch} by the results marked with the red diamond and star symbols.

Our findings allowed us to simplify the study by focusing only on sway for side-to-side motions and surge for fore-aft motions. We can extend the conclusions of these DoFs to pitch and roll motions, as long as the equivalent amplitude is below a certain value, which we estimate to be around $2^{\circ}$, i.e $A^* \approx 0.025$.

\subsection{Sway motion}
\label{sub-sec:side-to-side}

From the results presented in \S \S \ref{sub-section:equivalence_DOF} (figure \ref{fig:sway_roll}) we already see that the wake is strongly influenced by the $St$ (frequency of the motion). We observe that a higher frequency leads to a better wake recovery. To investigate the $St$ dependency, we measured the wake for different platform motion frequencies at constant amplitude ($A^* = 0.007$ and $St \in [0.12, 0.58]$) similar to the CFD simulations done by \cite{li2022onset}. 

\begin{figure}
  \centerline{\includegraphics[width=13cm]{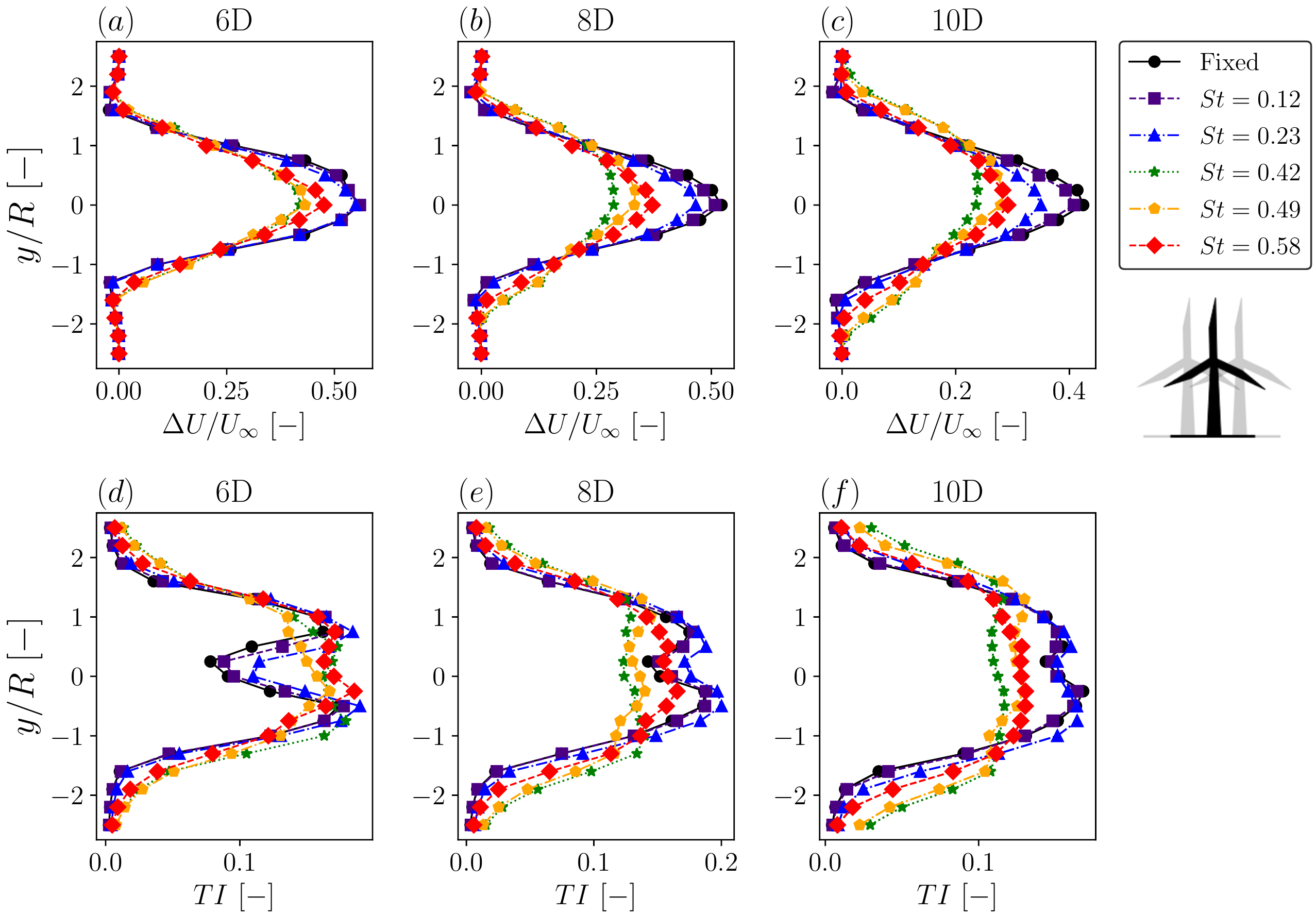}}
  \caption{Wake deficit (a-c) and $TI$ profiles (d-f) at 6D, 8D and 10D for fixed case and five \textbf{sway} cases with varying $St$ and constant $A^* = 0.007$, $\Rey = 2.3 \times 10^5$. Tests C.1-2 in table \ref{tab:cases_investigated}}
\label{fig:sway_st}
\end{figure}

\begin{figure}

  \centerline{\includegraphics[width=13.5cm]{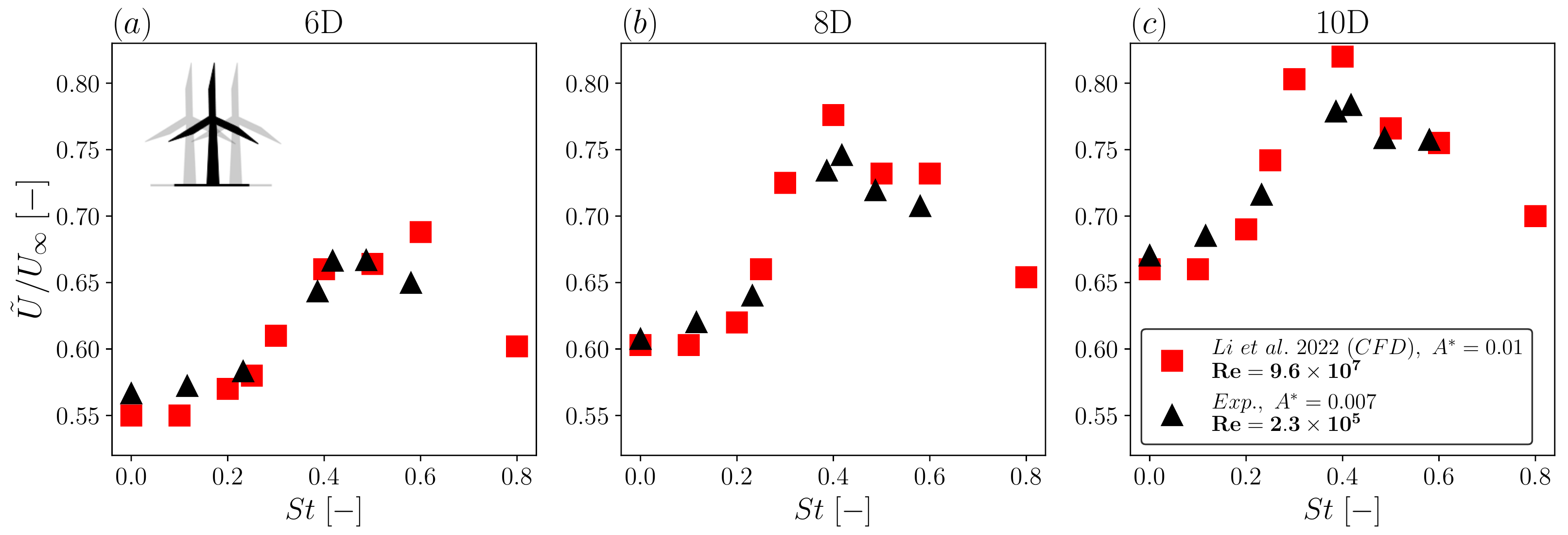}}
  \caption{Wake recovery expressed by the normalised average wind speed in the rotor area, $\tilde{U} / U_{\infty}$, defined in \S \S \ref{sub-sec:methodo} at 6D (a), 8D (b) and 10D (c) for fixed case and seven \textbf{sway} cases with varying $St$ and constant $A^* = 0.007$, $\Rey = 2.3 \times 10^5$. Tests C.1-2 in table \ref{tab:cases_investigated}. Also plotted with red squares: equivalent data from CFD simulations by \cite{li2022onset}, for which $A^* = 0.01$, $\Rey = 9.6 \times 10^7$ and $C_T$ unknown}
\label{fig:recov_sway_st}
\end{figure}

Figure \ref{fig:sway_st} shows the wake deficit and turbulence intensity ($TI$) profiles at 6D, 8D, and 10D for the fixed case and five sway cases. Most interestingly, we find that for $St = 0.42$ the wake has the lowest deficit (figure \ref{fig:sway_st} (b,c) green stars). Similar results are observed for the profiles of $TI$, see figure \ref{fig:sway_st} (e,f). Another interesting effect of wakes, discussed for instance by \cite{porte2020wind}, is the merging of the shear layers characterised by the vanishing of the two peaks in the profile of $TI$. In figure \ref{fig:sway_st} (d-f), it can be seen that the merging occurs for fixed and $St < 0.25$ between 8D and 10D. In contrast, the merging is found already around 6D for higher $St$.
\\
\\
To quantify the impact of motion frequency on the wind speed recovery, we computed $\tilde{U}(x)/U_{\infty}$, as explained in \S \S \ref{sub-sec:methodo}. Figure \ref{fig:recov_sway_st} shows our results together with the CFD simulations of \cite{li2022onset}. Although the two dataset are obtained for quite different $\Rey$, they match very well.  
We found that for each downstream position, the amount of wind recovered gradually increases up to a maximum for $St \approx 0.4$, after which it decreases. Sway movements of the turbine positively impact the recovery in the range of $St \in [0.3, 0.6]$, but the impact is less important outside of this range and approaches that of fixed case as $St \searrow 0$.
Comparing the results between $St = 0$ (fixed) and $St \approx 0.4$ (optimum) we found differences up to 25 \% in the recovery, this means that a wind turbine downstream could produced in this case ($St \approx 0.4$) significantly more power. 

\subsection{Surge motion}
\label{sub-sec:fore-aft-motion}

\begin{figure}
  \centerline{\includegraphics[width=13cm]{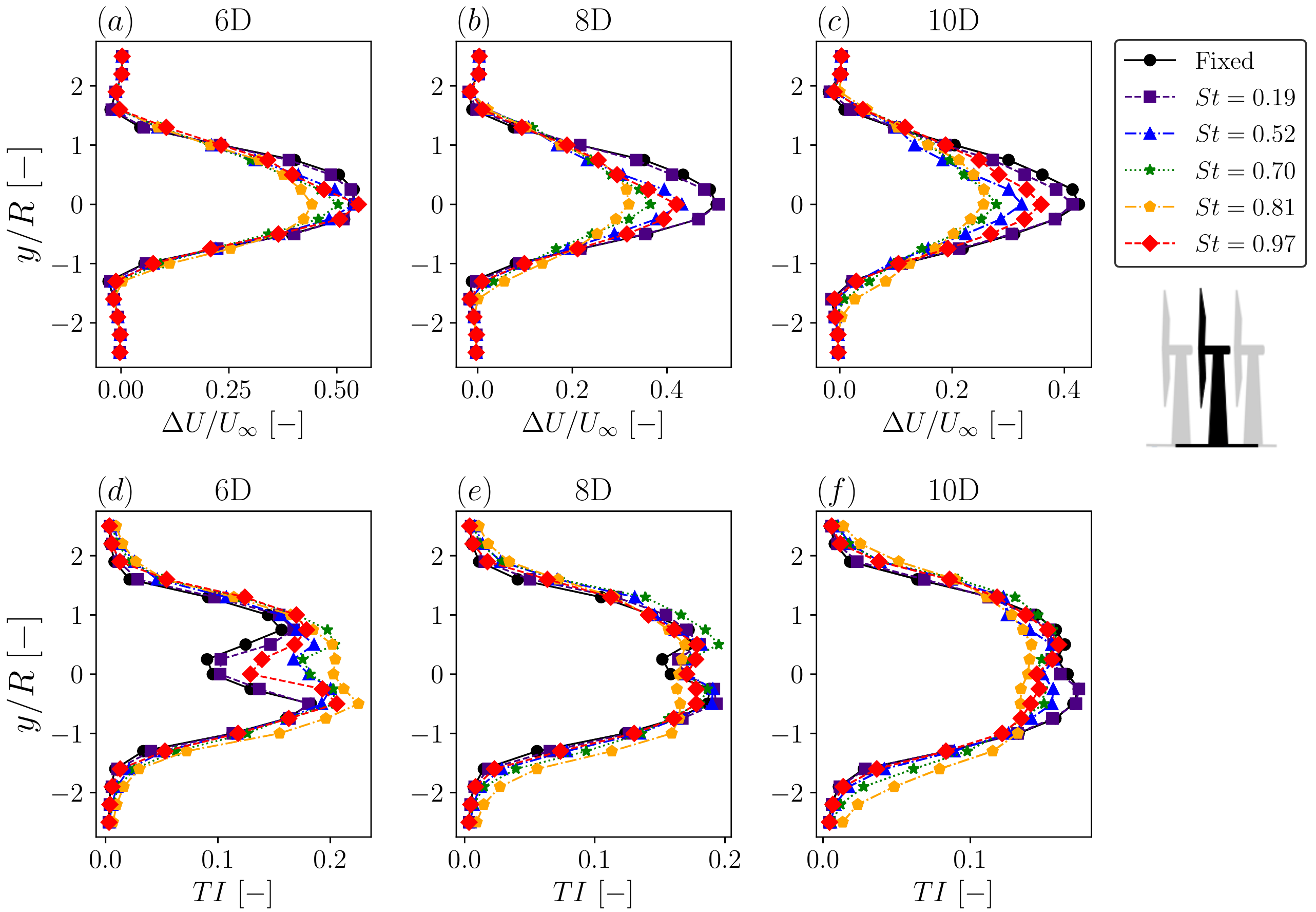}}
  \caption{Wake deficit (a-c) and $TI$ profiles (d-f) at 6D, 8D and 10D for fixed case and five \textbf{surge} cases with varying $St$ and constant $A^* = 0.007$, $\Rey = 1.4 \times 10^5$. Tests D.1-2 in table \ref{tab:cases_investigated}}
\label{fig:surge_st}
\end{figure}

We then present the results of the surge motions in a manner comparable to that of the sway. Based on a first experiment, we concluded about the necessity to investigate higher values of $St$ than for sway. In fact, we observed a plateau in the recovery for $St > 0.5$, which differs to the behaviour with sway motion (figure \ref{fig:recov_sway_st}). Thus we carried out experiments with $St \in [0, 0.97]$. To achieve this range of $St$, we used an inflow wind speed, $U_{\infty}$, of 3 m/s, see table \ref{tab:cases_investigated} (tests D.1-2).
\\
The wake deficit and $TI$ profiles are shown in figure \ref{fig:surge_st}. Notably, the wake for $St = 0.81$ has the lowest deficit, see profiles marked by yellow pentagons in figure \ref{fig:surge_st} (b,c). Likewise, the profile of $TI$ is the lowest at 10D for $St = 0.81$  (figure \ref{fig:surge_st} (f)). The merging of the shear layers occur at around 6D for $St \in [0.5, 0.8]$ (figure \ref{fig:surge_st} (d)) and between 8D and 10D for other cases.

\begin{figure}
  \centerline{\includegraphics[width=13.5cm]{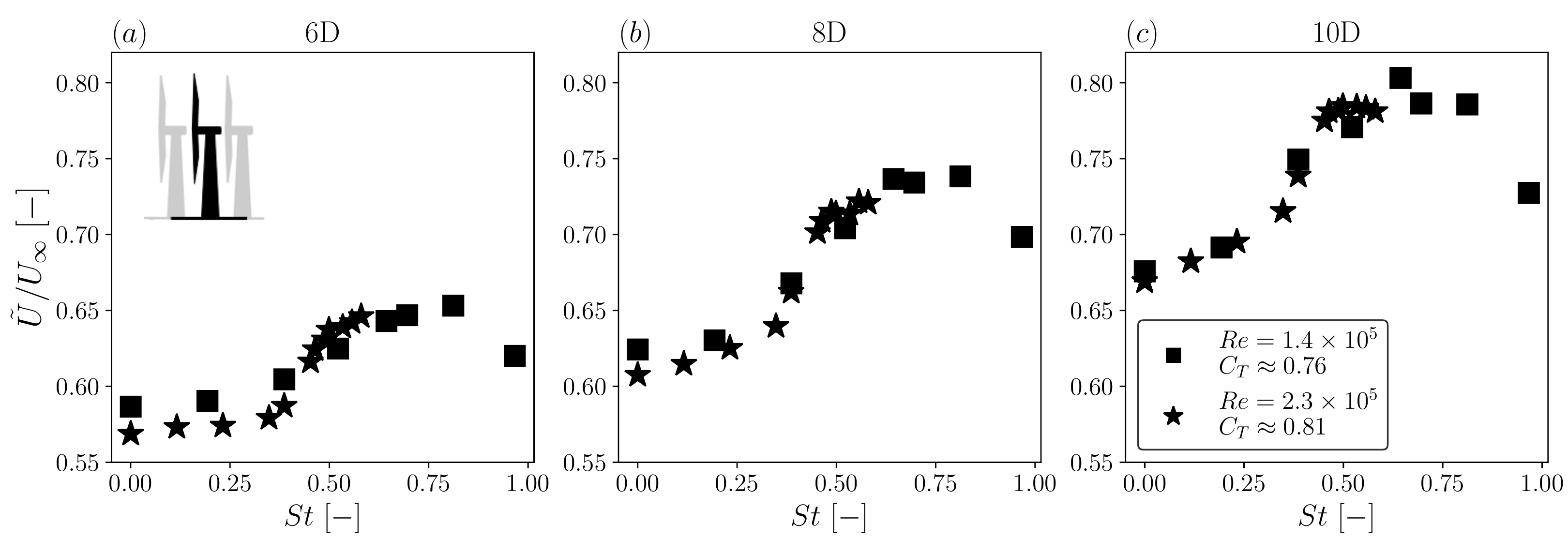}}
  \caption{Wake recovery expressed by the normalised average wind speed in the rotor area, $\tilde{U} / U_{\infty}$, defined in \S \S \ref{sub-sec:methodo} at 6D (a), 8D (b) and 10D (c) for fixed case and various \textbf{surge} cases with varying $St$, constant $A^* = 0.007$ and two $\Rey$: $1.4 \times 10^5,~ 2.3 \times 10^5$. Tests D.1-4 in table \ref{tab:cases_investigated}}
\label{fig:recov_surge_st}
\end{figure}

We also did the same experiments with $U_{\infty} = 5$ m/s, with which we could investigate $St$ up to 0.58 (D.3-4 in table \ref{tab:cases_investigated}). We thus have results for two $\Rey$, respectively $\Rey = 1.4 \times 10^5$ and $\Rey = 2.3 \times 10^5$. We calculated $\tilde{U}(x)/U_{\infty}$ (cf \S \S \ref{sub-sec:methodo}), to quantify the recovery with respect to $St$ as shown in figure \ref{fig:recov_surge_st}. For each position, the recovery increases for $St \in [0, 0.6]$, stays almost constant for $St \in [0.6, 0.8]$ and then decreases. Both $\Rey$ show comparable results. 

\subsection{Discussion}

The results show that, for both surge and sway motions, $St$ has a significant impact on wake recovery. Sway provides the highest wake recovery in the range of $St$ in $[0.3, 0.6]$ (figure \ref{fig:recov_sway_st}). For surge case, the high recovery range extends to higher $St$ values, namely in $[0.3, 0.9]$. Sway shows an optimal recovery well centred at $St^{opt} \approx 0.4$, whereas for surge the optimum is spread over a range of $St^{opt} \in [0.5, 0.8]$. Therefore, we conclude that the dynamical behaviour of the wake is likely to differ depending on the degree of freedom (DoF).
\\
Concerning the $TI$, the merging of the shear layers is an important property of the wake, which provides information about the wake development. We found that both types of motion cause an early merging (around 6D) compared to fixed case (between 8D and 10D), see $TI$ profiles in figures \ref{fig:sway_st} (d) and \ref{fig:surge_st} (d). Additionally, the maximum value of $TI$ is higher for the motion cases (up to 20 \% more than the fixed case). This indicates that movements generate extra turbulence in the form of small-scale turbulence as well as coherent structures, which accelerate the transition from near- to far-wake (as detailed later in the paper).
\\
Figure \ref{fig:recov_sway_st} shows the recovery for two $\Rey$ ($Re \sim 10^8$ from CFD and $Re \sim 2 \times 10^5$ in the experiments). Despite great differences in $\Rey$, the results match very well. For surge, we found that the recovery is similar for $\Rey = 1.4 \times 10^5$ and $\Rey = 2.3 \times 10^5$ (see figure \ref{fig:recov_surge_st}).  So we conclude that the wake of a floating turbine do not depend sensitively on $Re$, at least for $Re > 10^5$.
\\
The results of \S \ref{sec:results} motivated us for further investigations on the time resolved structures of the wakes, presented next in \S \ref{sec:discussion}.

\section{Wake dynamics}
\label{sec:discussion}

In the following \S \S \ref{sub-sec:sideways_discussion} and \S \S \ref{sub-sec:fore-aft_discussion}, we present the impact of the movements on the dynamics of the wake flow for sway and surge motions respectively. We then discuss these results in terms of non-linear dynamics in \S \S \ref{sub-sec:non-linear}.

\subsection{Wake dynamics under sway motion}
\label{sub-sec:sideways_discussion}

\begin{figure}
  \centerline{\includegraphics[width=13.5cm]{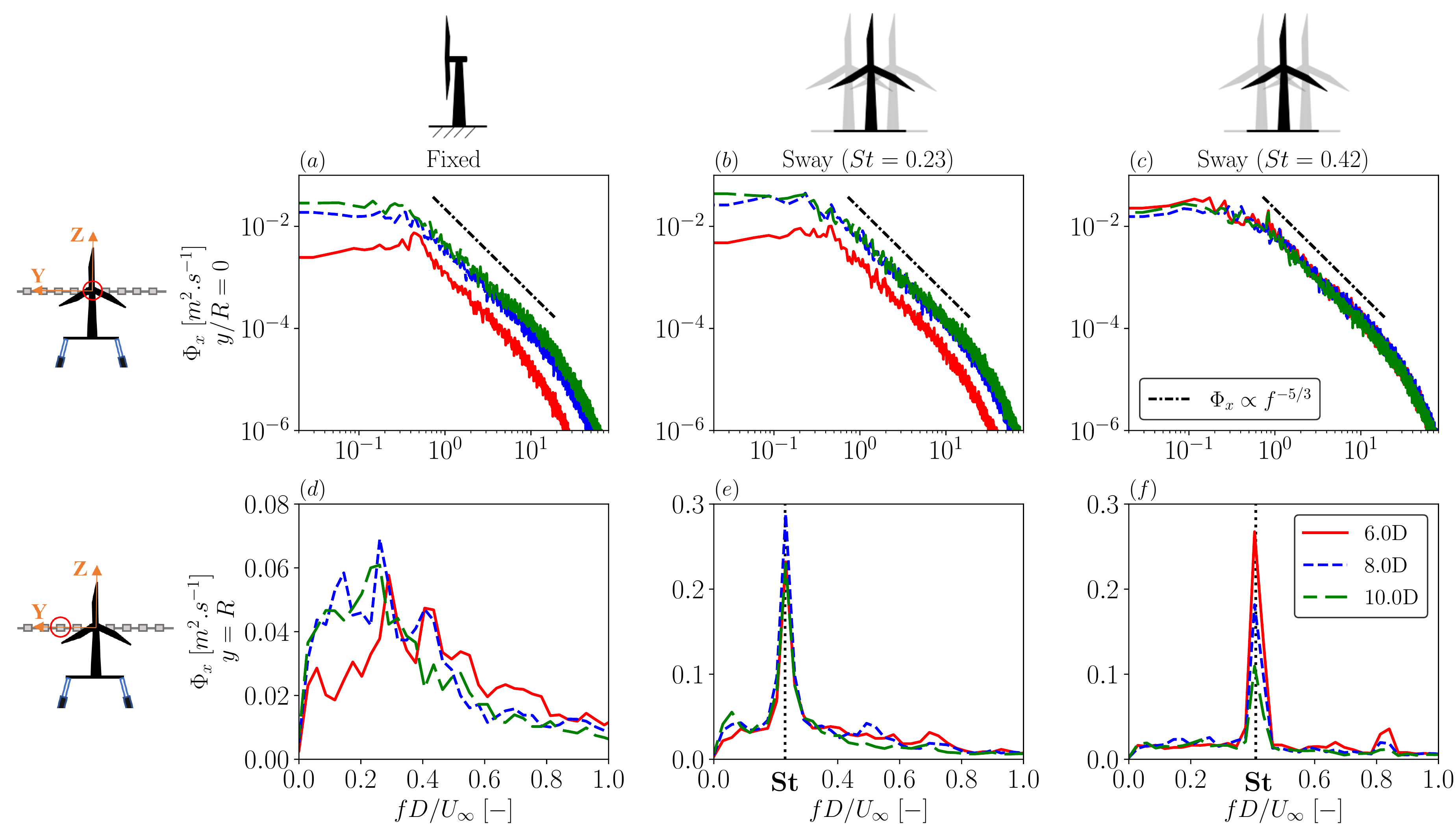}}
  \caption{Power spectra of the wind speed fluctuations in the wake at 6-10D for fixed case (a, d) and two \textbf{sway} cases ($St = 0.23$ (b, e) and $St = 0.42$ (c, f)) at two locations (y/R = 0 on top) and (y = R on bottom). $A^* = 0.007$, $\Rey = 2.3 \times 10^5$. Tests C.1-2 in table \ref{tab:cases_investigated} }
\label{fig:psd_sway}
\end{figure}

During a sway motion cycle, wakes are generated at various horizontal positions spanning $y \in [-A_p, +A_p]$. The mixing and interactions of these ``multiple" wakes affect the mean wake. For low frequencies, the wakes emitted at different horizontal positions superpose linearly (as shown in figure \ref{fig:sway_superposition} in appendix \ref{appA} and \cite{meng2022wind}). As in our case, we are interested in small amplitudes ($A^* \sim 0.01$) we assume that this linear superposition holds true if the profiles of $\Delta U / U_{\infty}$ and $TI$ coincide the fixed case. Clear deviations are interpreted as a non linear response. As seen in figure 6, for $St = 0.12$, the profiles match well the fixed case, but is not the case for higher $St$. 
\\
To further investigate the wake generated by the swaying turbine, we computed the power spectra, $\Phi_x$ (see \S \S \ref{sub-sec:methodo} for details) from the hot-wire measurements of the local velocity time series. Figure \ref{fig:psd_sway} displays $\Phi_x$ computed at $y = 0$ (a-c) and $y = R$ (d-f) for the following cases: fixed (a,d), $St = 0.23$ (b,e), $St = 0.42$ (c,f) for $x \in \llbracket 6D, 8D, 10D \rrbracket$.
\\
For $St = 0.42$, the spectra along the centreline all collapse for $fD/U_{\infty} > 1$, see figure \ref{fig:psd_sway} (c). In the inertia sub-range ($fD/U_{\infty} \in [1, \sim 20]$), $\Phi_x \propto f^{-5/3}$, which shows that turbulence in the wake center is fully developed \citep{pope2000turbulent, neunaber2020distinct}. Thus, the far-wake is reached at $x \leq 6D$ for $St = 0.42$, which is in line with the merging of the shear layers seen in figure \ref{fig:sway_st} (d). 
\\
For $St = 0.23$, the spectra at $x = 8D$ and $x = 10D$ merged, indicating that the far-wake region is reached between 6D and 8D. For the fixed case, all three spectra are unequal, showing that the far-wake first appears at $x > 8D$.

\begin{figure}
  \centerline{\includegraphics[width=15cm]{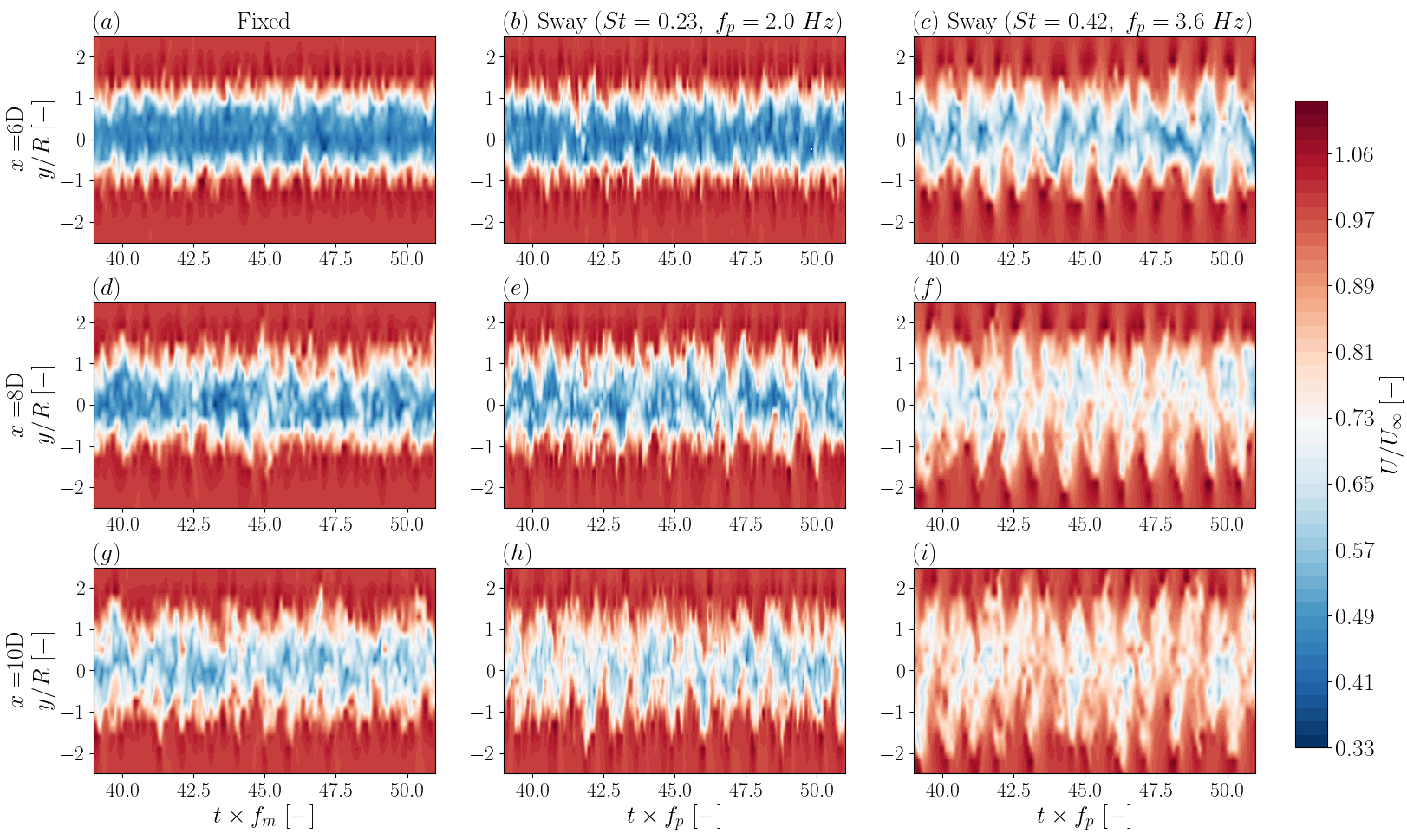}}
  \caption{Instantaneous wind field in the wake at 6-10D for fixed case (a,d,g) and two \textbf{sway} cases ($St = 0.23$ (b,e,h) and $St = 0.42$ (c,f,i)). $A^* = 0.007$, $\Rey = 2.3 \times 10^5$. Tests C.1-2 in table \ref{tab:cases_investigated}. \textit{For fixed case, time is multiplied by $f_m = 0.35 \times U_{\infty}/D$}. \textit{For moving cases, time is multiplied by $f_p$}.}
\label{fig:sway_meandering}
\end{figure}

At $y = R$, the spectra of the fixed turbine display a region of high turbulent energy for $fD/U_{\infty} \in [0.1, 0.5]$, showing the presence of flow structures with characteristic frequencies in this range (figure \ref{fig:psd_sway} (d)). At $x = 6D$, the highest energy is contained in $f_m D/U_{\infty} \approx 0.35$. 
According to \cite{foti2018similarity}, the far-wake of a wind turbine experiences meandering at a natural frequency, typically in the range of $f_m \in [0.1,0.5]U_{\infty}/D$. \cite{gupta2019low} further explains that the inherently erratic wake tends to amplify small perturbations, resulting in wake meandering far downstream. The broad peak at $f_m D/U_{\infty}$ is consistent with this explanation, suggesting that such meanderings result from shear flow instabilities. These instabilities span a range of frequencies and do not show a clear peak, as would be typical for vortex shedding.
\\
The spectra of moving cases show a distinct peak at the frequency of movements for all downstream positions, see figure \ref{fig:psd_sway} (e,f). This peak is a signature of the motions within the wake, indicating that the far-wake contains coherent structures at the frequency of the motion. This was also observed by \cite{fu2019wake} and \cite{li2022onset}. This pseudo lock-in phenomenon, as described in \cite{gupta2019low}, occurs when the wake flow synchronises with the forcing frequency imposed by the periodic upstream perturbation (in our case the movements of the platform). Pseudo lock-in is observed when the platform's motion frequency, $f_p$, is around $f_m$, and seems to be the strongest when $f_p$ is closest to $f_m$. The intensity of the lock-in is highest at $x=6D$ for $St=0.42$. For other cases, the highest lock-in occurs at $x=8D$, indicating a spatial dependency. From the spectra analysis, we can conclude that the wake dynamics are closely following the motions frequency.
\\
The array of hot-wires, with 19 probes aligned horizontally, enables visualisation of the instantaneous wake flow of the turbine (see \S \S \ref{sub-sec:methodo} for details). 
Figure \ref{fig:sway_meandering} shows the time evolution of $U(x,y,t)$ for the fixed case (a,d,g), $St = 0.23$ (b,e,h), and $St = 0.42$ (c,f,i) at the three downstream positions. For the fixed case, time (x-axis) is multiplied by $f_m \approx 0.35 U_{\infty}/D$. In the case of the moving turbine, time is multiplied by the frequency of motion of the platform. 
\\
For the above mentioned case of the highest lock-in response ($St = 0.42$ at $x = 6D$) we see correspondingly a clear meandering pattern at the imposed frequency, $f_p$ (figure \ref{fig:sway_meandering} (c)). As we move downstream, the amplitude of meandering increase and the structures become more fuzzier.
\\
For lower $St$ of 0.23, we see how the structures seem to be a combination of the flows of the fixed case and $St = 0.42$. For fixed and $St = 0.23$, the meandering structures become more prominent for larger distances.
\\
Overall, the clearest meandering  structures are obtained for $St = 0.42$ quite close to the rotor ($x = 6D$). Whereas the other cases have a tendency to build up less clear meandering structures only further downstream.  
\\
\\
Coming back to the profiles of wake deficit, we found that the wake expansion is the largest for $St = 0.42$ (figure \ref{fig:sway_st} (b, c)), which coincides with the biggest meandering amplitudes of the wake field (figure \ref{fig:sway_meandering} (c,f,i)). In figure \ref{fig:recov_sway_st}, we define the wake recovery over a range of $y \in \sim[-R, R]$, which corresponds to the energy available for a virtual turbine operating in this wake. If a larger range is considered, the recovery is less significant but still higher.
From our time-resolved measurements, we find that, compared to the fixed case, the swaying motions of the platform enhance the sideways motion of the wake, which distributes the momentum more evenly within $y \in [-2.5R, 2.5R]$. The larger surface area of the wake as well its dynamics enable more transport of momentum to the center of the wake. 

\begin{figure}
  \centerline{\includegraphics[width=15cm]{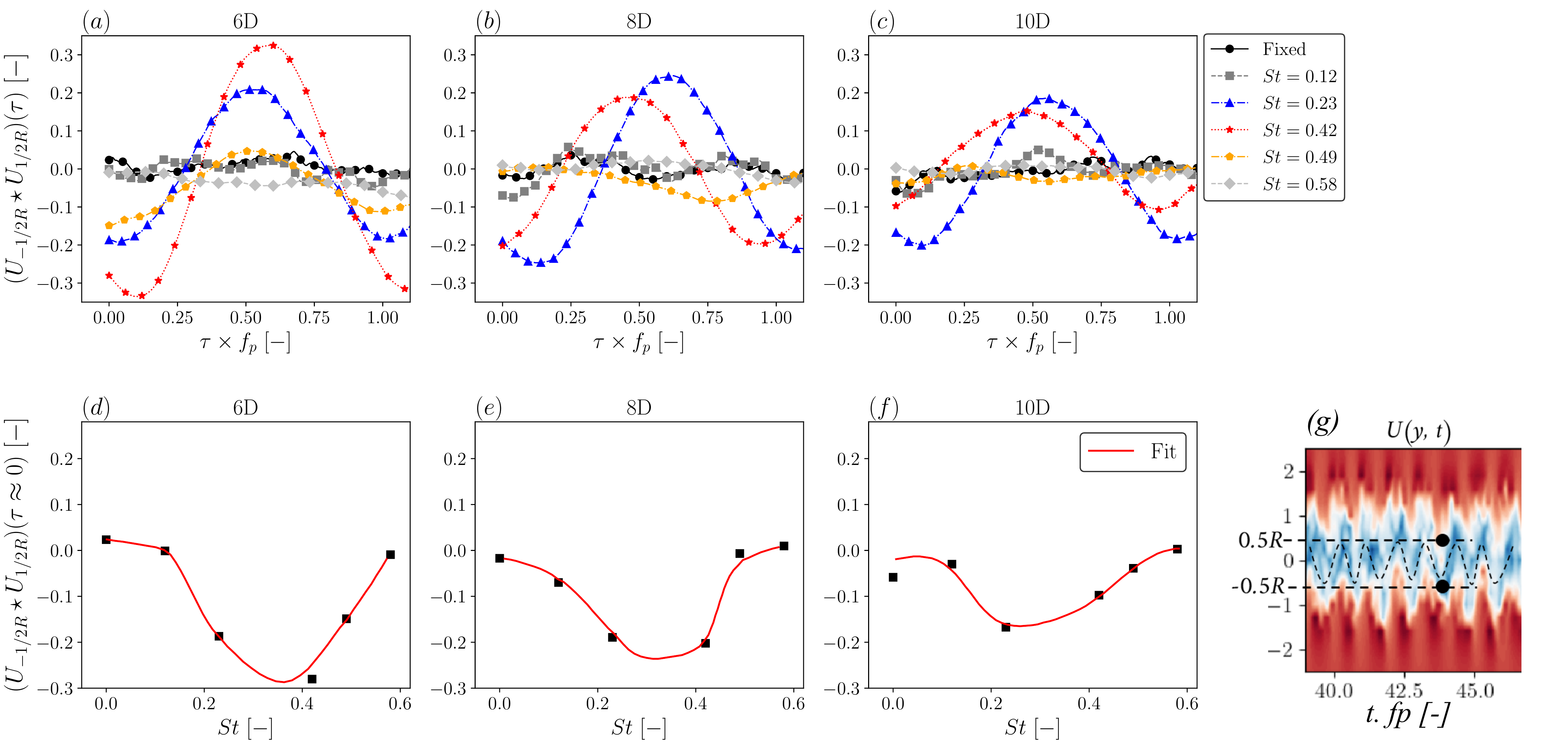}}
  \caption{Cross-correlation between $U(x,-1/2R,t)$ and $U(x,1/2R,t+\tau)$, noted $(U_{-1/2R} \star U_{1/2R}) (\tau)$,  for fixed case and \textbf{sway} cases with varying $St$ at 6-10D (a,b,c). Plot of  $(U_{-1/2R} \star U_{1/2R}) (\tau \approx 0)$ vs. $St$ (d-f). $A^* = 0.007, ~\Rey = 2.3 \times 10^5$. Tests C.1-2 in table \ref{tab:cases_investigated}. \textit{ (g) is an illustration of the cross-correlation for $\tau = 0$}.}
\label{fig:sway_correlation}
\end{figure}

The spectra in figure \ref{fig:psd_sway} is not a complete characterisation of a meandering wake. However, the instantaneous wake flow fields in figure \ref{fig:sway_meandering} clearly indicate meandering. To further quantify the meandering at a given frequency, we calculated the cross-correlation function (depicted in \S \ref{sub-sec:methodo}) between $U(x,-1/2R,t)$ and $U(x,1/2R,t+\tau)$ for $\tau$ varying from 0 to $3/f_p$. Figure \ref{fig:sway_correlation} shows this quantity computed at 6D (a), 8D (b), and 10D (c) for both fixed and a few swaying cases. We plot $(U_{-1/2R} \star U_{1/2R}) (\tau)$ versus $\tau \times f_p$, except for the fixed case where we set $\tau \times f_p = \tau$. In figure \ref{fig:sway_correlation} (a), it can be seen that for $St \in [0.2,0.5]$, $(U_{-1/2R} \star U_{1/2R}) (\tau)$ is a harmonic function with a period equal to the platform's oscillation period, $1/f_p$, consistent with the pseudo lock-in phenomenon. The function attains its minimum value (negative value) at $\tau \approx 0$. For $x = 6D$ and $St = 0.42$, the minimum value of $(U_{-1/2R} \star U_{1/2R}) (\tau \approx 0)$ is approximately $-0.3$, which indicates that the signals $U(-1/2R,t)$ and $U(1/2R,t)$ are anti-correlated. This anti-correlation quantifies the side-to-side motion of the wake at a specific frequency. When the wake meanders to the left, $U(1/2R,t)$ decreases and $U(-1/2R,t)$ increases and vice versa (for illustration see figure \ref{fig:sway_correlation} (g)). Consistently, the correlation is maximal at $\tau \approx 1/2f_p$.

In figure \ref{fig:sway_correlation} (d-f), $(U_{-1/2R} \star U_{1/2R}) (\tau \approx 0)$ is plotted vs. $St$ for the three downstream positions. In figure \ref{fig:sway_correlation} (d), the anti-correlation is significant in the range of $St \in [0.2, 0.5]$ and close to zero for other values, confirming the range of motion frequencies that initiate large wake meandering.

\subsection{Wake dynamics under surge motion}
\label{sub-sec:fore-aft_discussion}

\begin{figure}
  \centerline{\includegraphics[width=14cm]{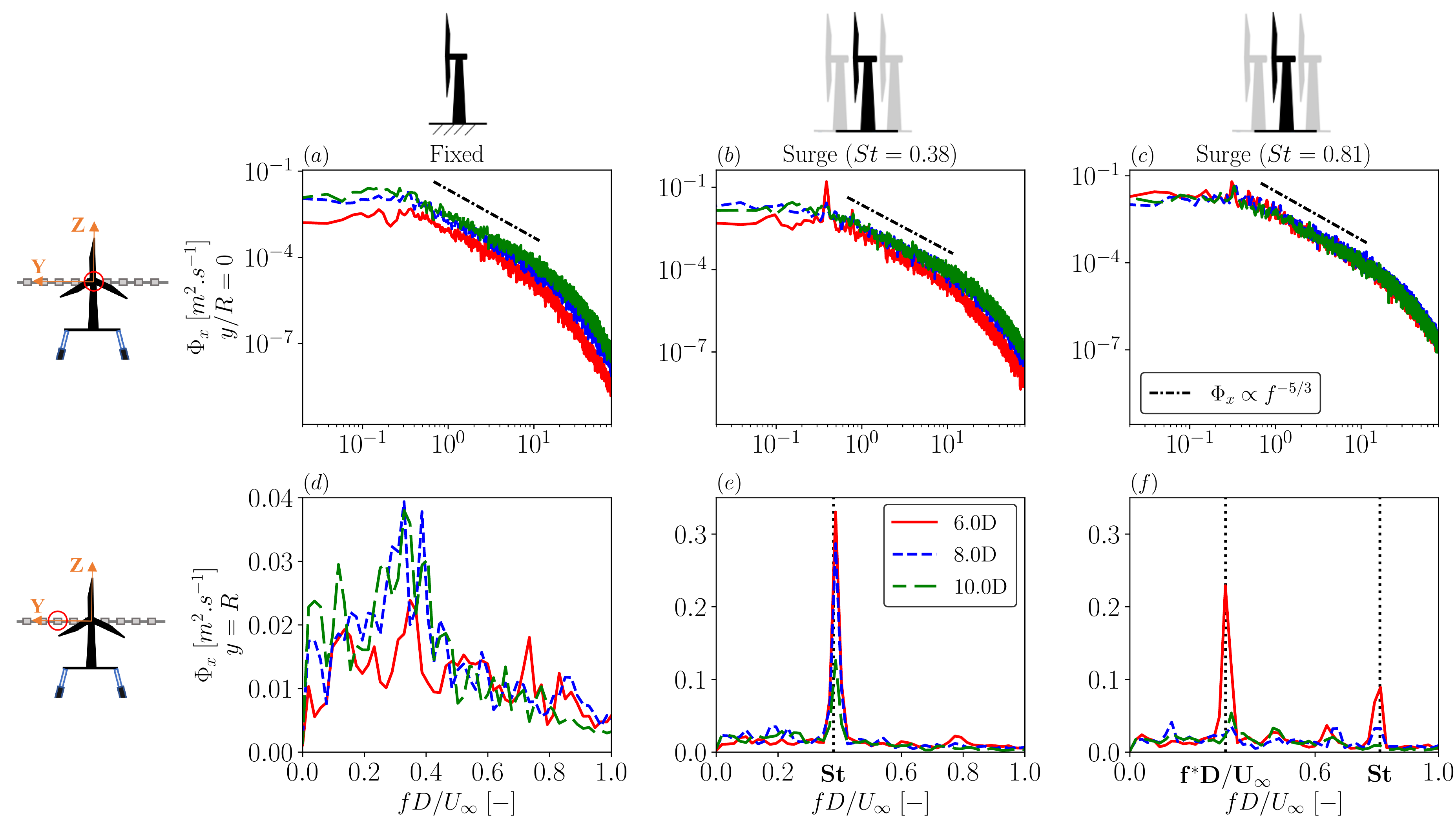}}
  \caption{Power spectra of the wind speed fluctuations in the wake at 6-10D for fixed case (a, d) and two \textbf{surge} cases ($St = 0.38$ (b, e) and $St = 0.81$ (c, f)) at two locations (y/R = 0 on top) and (y = R on bottom). $A^* = 0.007$, $\Rey = 1.4 \times 10^5$. Tests D.1-2 in table \ref{tab:cases_investigated} }
\label{fig:psd_surge}
\end{figure}

Surge motions, unlike sway motions, induce variations of $C_T$ that affect the dynamics of the overall wake. $C_T$ oscillates at $f_p$ and is out of phase of around $\pi/2$ with the signal of motions as shown by \cite{fontanella2021unaflow}. So $C_T (t) = <C_T> + \Delta C_T sin(2 \pi f_p t + \phi)$, with $\Delta C_T$ the amplitude of thrust oscillation and $\phi$ the phase shift with the signal of motion. In our case, the highest value of thrust coefficient variation is:
 $\Delta C^{max}_T \approx 0.07$ for $f_p = 5$ Hz,$~A_p$ = 0.004 m,$~U_{\infty} = 3$ m/s. Besides the $C_T$ variations, the wake is generated behind the turbine at various $x$ locations spanning $[-A_p, +A_p]$ during one cycle of motion.
\\
As for sway, when $St < 0.1$, the wake of the surging turbine is similar to that of the fixed turbine, even at high amplitude (see figure \ref{fig:surge_pitch} (c)), which is consistent with \cite{schliffke2020wind, meng2022wind, belvasi2022far}. When the frequency is higher, $St > 0.2$, the interactions of the wakes emitted at different $x$ positions and with different $C_T$ become more complex. 
\\
\\
To understand the impact of surge movements on the dynamics of the wake we carried out a similar analysis to that made for sway (\S \S \ref{sub-sec:sideways_discussion}). Figure \ref{fig:psd_surge} shows the power spectra of fixed (a,d), surge with $St = 0.38$ (b,e) and surge with $St = 0.81$ (c,f) for the three downstream positions at $y = 0$ (a-c) and $y = R$ (d-f). 
\\
The spectra of $St = 0.81$ at $y = 0$ all collapsed to one spectra for $fD/U_{\infty} > 1.0$, and $\Phi_x \propto f^{-5/3}$ (for $fD/U_{\infty} \in [1, \sim 20]$) which confirms that the far-wake is reached even for $x \leq 6D$. Whereas for the lowest $St$ and the fixed case, the the far-wake is developed for larger distances, $x > 6D$. This behaviour is very similar to that observed for the swaying turbine. 
\\
$\Phi_x$ at $y = R$ exhibit a pronounced peak at the motion frequency ($St$) for both surge cases, in line with results from \cite{schliffke2020wind, belvasi2022far}. As for sway, this indicates that the wake of the surging turbine contains coherent flow structures with a characteristic frequency of $f_p$. The peak is for both $St$ maximum at $x = 6D$ and then decreases.
\\
When looking at the spectra of the fixed turbine at $y = R$, the frequency range with the largest amount of energy is similar ($0.1 < fD/U_{\infty}< 0.5$) to that of the other fixed case in figure \ref{fig:psd_sway} (d). The maximum energy is located at $f_m \approx 0.3 U_{\infty}/D$, which is close to than in figure \ref{fig:psd_sway} (d). These two cases have a $\Rey$ from $2.3 \times 10^5$ in figure \ref{fig:psd_sway} to $1.4 \times 10^5$ in figure \ref{fig:psd_surge}.
\\
Looking at the case of $St = 0.38$, we see interestingly that all frequency components of the fixed case merge to one narrow peak corresponding to the excitation frequency with $St = 0.38$.  This is somehow similar to the sway cases (see figure \ref{fig:psd_sway} (e,f)). For $St = 0.81$, compared to the sway case, a new dynamical behaviour is observed, namely we find two frequencies in the spectra with apparent mixing components. One peak is at the excitation frequency, i.e $St = 0.81$. The other peak at $0.31U_{\infty} /D$ seems to be due to a self generated mode, $f^*$ of the wake (around $f_m$, the meandering frequency of the fixed turbine). Most interestingly, the further smaller peaks are related to linear combinations of the two frequencies, i.e with $a(f^*D/U_{\infty}) + b St$. One small peak is around (a,b) = (1,-1/2) and another one at (a,b) = (1,-1). This proves non-linear mode coupling, which is discussed later in \S \S \ref{sub-sec:non-linear}. For other values of $St \in [0.6, 0.9]$, we also observe a second large peak at a frequency, $f^*$ in $[0.15,0.35]U_{\infty} /D$ like the one at $0.31U_{\infty} /D$ for $St = 0.81$. It is important to note that the value of $f^*$ changes with the exciting frequency, $St$.

\begin{figure}
  \centerline{\includegraphics[width=15cm]{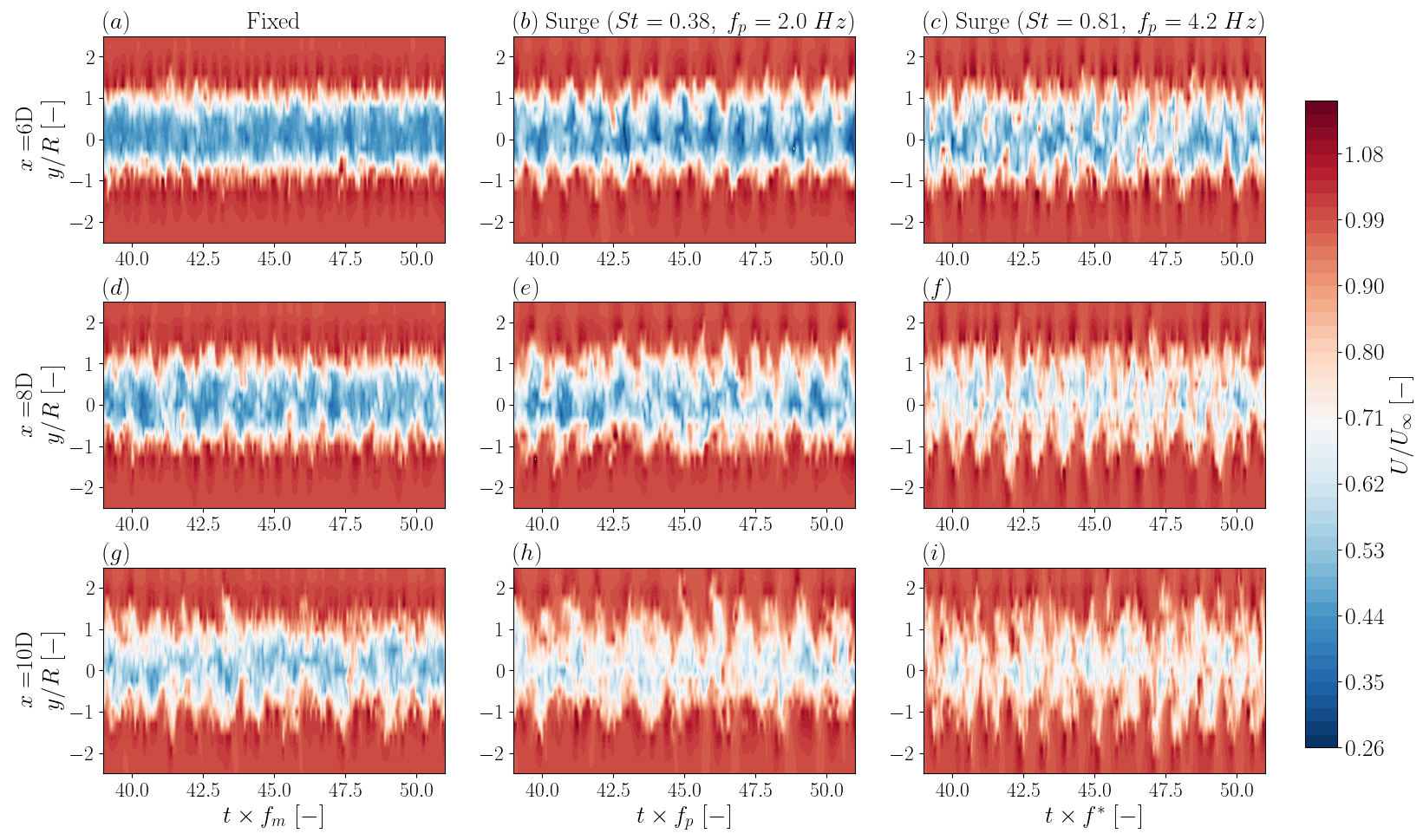}}
  \caption{Instantaneous wind field in the wake at 6-10D for fixed case (a,d,g) and two \textbf{surge} cases ($St = 0.38$ (b,e,h) and $St = 0.81$ (c,f,i)). $A^* = 0.007$, $\Rey = 1.4 \times 10^5$. Tests D.1-2 in table \ref{tab:cases_investigated}. \textit{For fixed case, time is multiplied by $f_m \approx 0.30 U_{\infty}/D$. For $St = 0.38$, time is multiplied by $f_p$ and for $St = 0.81$ is multiplied by $f^* \approx 0.31 U_{\infty}/D$}.}
\label{fig:surge_meandering}
\end{figure}

As for sway, we are interested in the spatio-temporal structure of these oscillating modes. We show in figure \ref{fig:surge_meandering} the instantaneous wake flows field for the three cases: fixed case (a,d,g), $St = 0.38$ (b,e,h), and $St = 0.81$ (c,f,i). 
\\
The most striking flow pattern is seen at $x = 6D$ for $St = 0.38$ (figure \ref{fig:surge_meandering} (b)). The wake is pulsating at the frequency of the motions. In the wake of the two further downstream positions (figure \ref{fig:surge_meandering} (e, h)), the pulsing is gradually reduced. 
\\
The flow dynamics at $St = 0.81$ are quite different, the wake at 6D (figure \ref{fig:surge_meandering} (c)) is clearly meandering with a frequency of $f^* \approx 0.31U_{\infty}/D$, in consistency with the spectra of this case (figure \ref{fig:psd_surge} (f)). Further downstream, the periodic structures become more irregular, which is also seen in the reduced peaks in the spectra at $x = 8D$ and $x = 10D$.

\begin{figure}
  \centerline{\includegraphics[width=14.5cm]{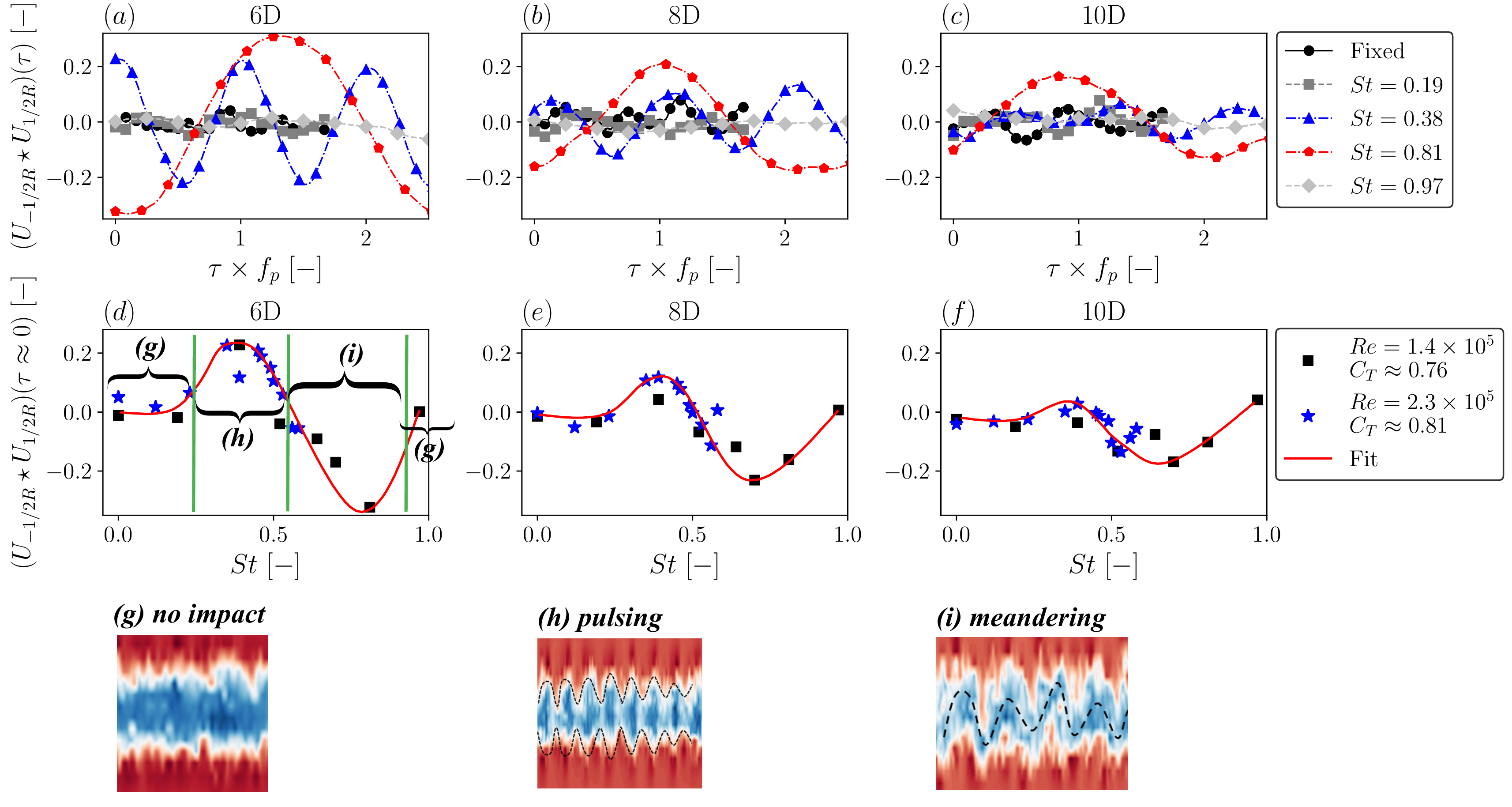}}
  \caption{Cross-correlation between $U(x,-1/2R,t)$ and $U(x,1/2R,t+\tau)$, noted $(U_{-1/2R} \star U_{1/2R}) (\tau)$,  for fixed case and \textbf{surge} cases with varying $St$ at 6-10D (a,b,c). Plot of  $(U_{-1/2R} \star U_{1/2R}) (\tau \approx 0)$ vs. $St$ (d-f). $A^* = 0.007, ~\Rey = 1.4 \times 10^5$ and $\Rey = 2.3 \times 10^5$. Tests D.1-4 in table \ref{tab:cases_investigated}. \textit{Picture (g) depicts the wake of the fixed turbine, picture (h) shows a typical pulsating motion of the wake and picture (i) displays a typical meandering pattern} }
\label{fig:surge_correlation}
\end{figure}

To investigate the pulsing and meandering modes of the wake in more details, we computed the cross-correlation between the wind speed signals $U(x,-1/2R,t)$ and $U(x,1/2R,t+\tau)$ (cf \S \ref{sub-sec:methodo}). We calculated $(U_{-1/2R} \star U_{1/2R}) (\tau)$ for surge cases E.1-4 (see table \ref{tab:cases_investigated}). Figure \ref{fig:surge_correlation} shows $(U_{-1/2R} \star U_{1/2R}) (\tau)$ as a function of $\tau \times f_p$. In figure \ref{fig:surge_correlation} (a), we observe that the cross-correlation is almost zero everywhere for $St < 0.25$ and $St \approx 1.0$, indicating that the wake does not oscillate at a clear frequency. 
\\
For $0.25 < St < 0.55$, $(U_{-1/2R} \star U_{1/2R}) (\tau)$ is a harmonic function with a period of $1/f_p$, and the cross-correlation is maximal (positive) at $\tau \approx 0$. The velocity fluctuations between $y=-1/2R$ and $y=1/2R$ are in phase, which implies that the wake undergoes pulsating movements and not meandering.
\\
On the other hand, for $St = 0.81$, $(U_{-1/2R} \star U_{1/2R}) (\tau)$ is a harmonic function with a period of $\sim D/(0.31 U_{\infty})$, and the cross-correlation is minimum (negative) at $\tau \approx 0$. This behaviour is associated with meandering movements of the wake at a given frequency, as discussed for sway in \S \ref{sub-sec:sideways_discussion} and is consistent with the instantaneous flow fields shown in figure \ref{fig:surge_meandering} (c, f, i). 
\\
Next we show that the analysis of the cross-correlation functions allows to distinguish the different ranges of $St$ for which the wake's dynamic behaviour changes. In figure \ref{fig:surge_correlation} (d-f) the $St$ dependency of the cross-correlation at $\tau \approx 0$ is shown. The fit of this cross-correlation allows to identify three distinct regions, highlighted in figure \ref{fig:sway_correlation} (d) and schematically visualised in figure \ref{fig:sway_correlation} (g-i). For $St < 0.25$ and $St \approx 1.0$ (zone (g)), the motions do not significantly impact the wake's dynamic, which is similar to that of the fixed turbine (as illustrated in figure \ref{fig:surge_correlation} (g)). For $St \in [0.25, 0.55]$ (zone (h)), the positive correlation values indicate that the wake undergoes pulsing movements (maximum correlation at $St \approx 0.4$). Picture (h) shows typical pulsating pattern in the wake. For $St > 0.55$, $(U_{-1/2R} \star U_{1/2R}) (\tau \approx 0)$ decreases and becomes negative, indicating that the wake meanders, as shown in picture (i). 
\\
Further downstream, our analysis indicates that pulsating motions as well as the meandering motions gradually vanish. In figure \ref{fig:surge_correlation} (e,f), the cross-correlation values are overall reduced. 

\subsection{Discussion in terms of non-linear dynamical system}
\label{sub-sec:non-linear}

\begin{figure}
  \centerline{\includegraphics[width=15cm]{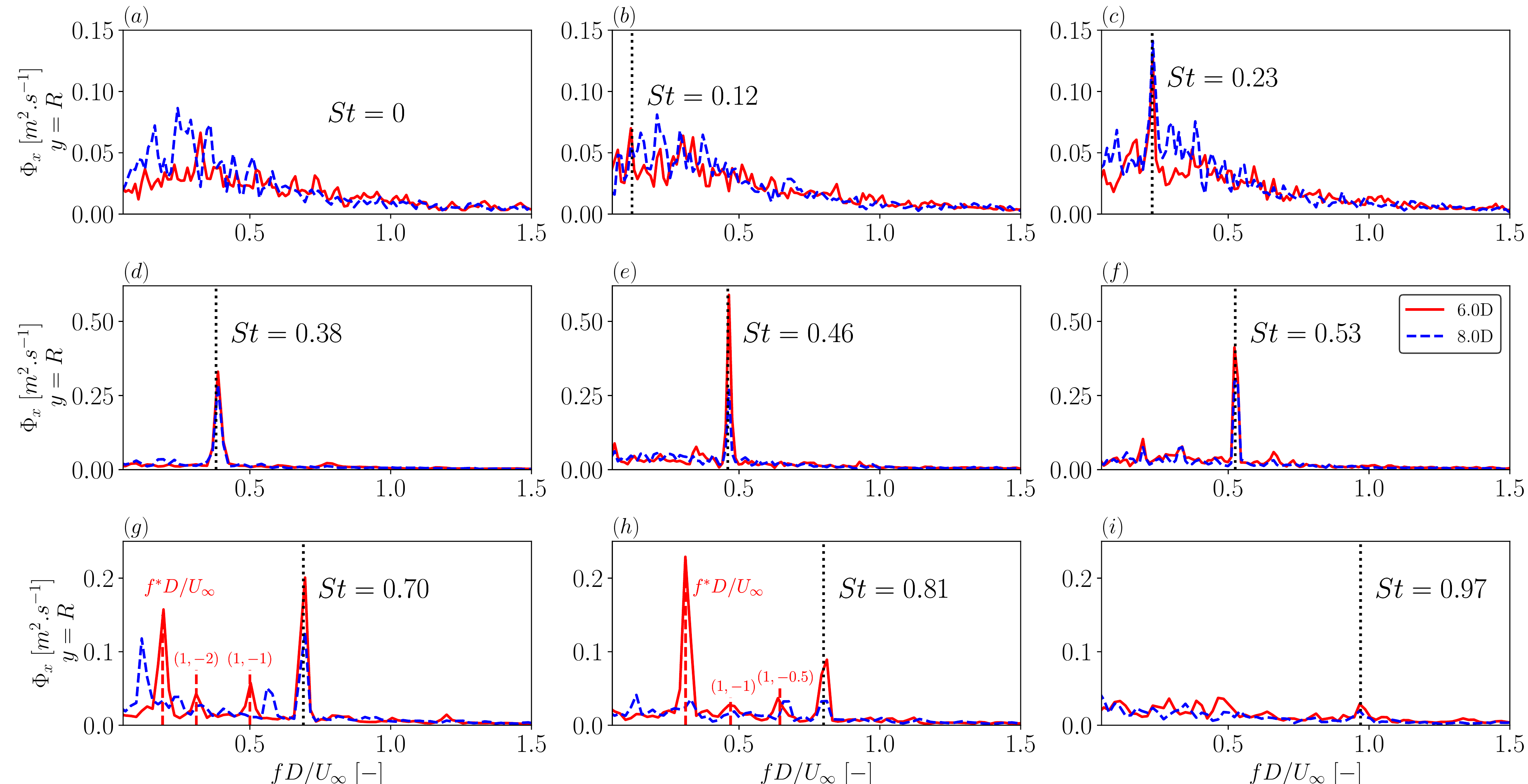}}
  \caption{Power spectra of the wind speed fluctuations in the wake at $x=6D$ and $x=8D$ for fixed case (a) and 8 \textbf{surge} cases (b-i) at the location y = R. $A^* = 0.007$, $\Rey = 1.4$ to $2.3 \times 10^5$. Tests E.1-4 in table \ref{tab:cases_investigated} }
\label{fig:psd_non_linear}
\end{figure}

The following discussion concentrates on the boundary region of the wake which is dominated by shear ($y \approx R$). First, we summarise the observations made for the wake  of the fixed turbine. For this case, we found a kind of broadband coloured noise in the range of $fD/U_{\infty} \in  [0.1, 0.5]$ (see figure \ref{fig:psd_sway} (d) and figure \ref{fig:psd_surge} (d)) with a maximum at about $0.35$ which we denoted as meandering frequency, $f_m$. As mentioned in the introduction, this broadband noise is consistent with  \cite{okulov2014regular, foti2018similarity, heisel2018spectral, gupta2019low}. These dynamics are most likely the result of the development of shear layer instabilities.
\\
\\
Clear differences in wake dynamics are seen between sway and surge motions. Whereas under sway movements, one clear wake meandering mode is found, we observe two modes for surge, namely a pulsing (for $St \in [0.25, 0.55]$) and a meandering mode (for $St \in [0.55, 0.9]$) as seen in figure \ref{fig:surge_correlation} (d). The pulsing mode vanishes, or gets damped out, for larger distances ($x > 8D$) as well as for higher $St$ numbers ($St > 0.6$). Adjacent to this pulsing mode region of low distances and low $St$, a smeared-out (turbulent) meandering mode is observed (see figure \ref{fig:surge_meandering} (b) for clear pulsing, (h) for smeared out pulsing, (c) for clear meandering mode and (i) for more ``turbulent" meandering).  
\\
\\
These modes can be interpreted in terms of general characteristics of non-linear dynamics \citep{peinke2012encounter, argyris2015exploration}. The first remarkable feature is the synchronisation of the wake dynamic to the forcing frequency of the platform motion (see figure \ref{fig:psd_sway} (f) for sway and figure \ref{fig:psd_surge} (e) for surge). Compared to the spectra of the fixed case, we clearly see how the broad band frequency gets slaved to a narrow band peak \citep{haken2012advanced}. Such a synchronisation effect, or pseudo lock-in, is a prominent feature of non-linear dynamics discussed in various contexts. It is well known that many coupled pendulums with different frequencies have the tendency to synchronise to one common frequency \citep{acebron2005kuramoto}. If so, this suggests that the synchronisation effect is caused by the driving surge or sway motion, which has a similar effect to that of the coupling of the pendulums in the case mentioned above. Another explanation may be given by the so-called stochastic resonance, which describes how a noisy system gets into into resonance with a small exciting periodic perturbation \citep{benzi1981mechanism}.
This is consistent with our observations that the amplitude of the meandering and pulsing due to platform movements ($\sim 0.01D$) is about one to two orders of magnitude larger ($\sim D$) than the platform amplitude, indicating a significant amplification of the small disturbances that grow downstream. Another characteristic of synchronisation is that it is robust to changes in the driving frequency within a finite range. At the end of such a synchronisation range, other non-linear effects emerge. The tracking to the driving frequency is clearly seen for our experiments (see figure \ref{fig:psd_non_linear} (c-f)). 
In summary, we observe a synchronisation of the dynamics of the boundary region of the wake to the platform motion together with a high amplification of the amplitude. A strong reduction of the broad band structure of the initial frequency (wake of the fixed turbine) and a locking to the exciting frequency is clearly observed for $St \in [0.25, 0.55]$ for both sway and surge. This effect corresponds to a significant reduction in the wake dynamical degrees of freedom, in the sense that the non-linear dynamic oscillation has a low dimension, with only two degrees of freedom.
\\
\\
In contrast to sway, the surge motion shows further phenomena of non-linear dynamics. In fact, two spatio temporal wake patterns are observed, namely the pulsing and the meandering mode (figure \ref{fig:surge_correlation} (h, i)). As discussed above, for surge with $St >0.55$, two frequencies with mixing components are seen in the spectra (figure \ref{fig:psd_surge} (f)). This indicates that the wake synchronises to two frequencies, namely the exciting frequency, $f_p$ and a self-generated narrow band mode, $f^*$. A coupling of these frequencies is a clear sign of non-linear dynamics. In figure \ref{fig:psd_non_linear}, we show the psd at $y = R$ and $x = 6D$  and $x=8D$ for a few surge cases with $St \in [0, 0.97]$. Very interesting are the spectra of $St = 0.7$ (g) and $St = 0.81$ (h). We see a peak at $St$ and at $f^*$ (the self generated mode) as well as mixing components in the form of $a(f^*D/U_{\infty}) + b St$. For $St = 0.70$, for instance, we identify two further peaks with (a,b) = (1,-1) and (a,b) = (1,-2). Such a dynamic is part of the class of quasi-periodic systems, described with the generic model of circle map, which characterises the quite complex and fast changing dynamical behaviour under such non-linearities \citep{argyris2015exploration}. We observe that one time the amplitude of the driving frequency is larger and the other time the amplitude of the self generated mode is greater (see figure \ref{fig:psd_non_linear} (g) $x=6D$ and $x=8D$). Mixing components occur more or less pronounced, which is similar to previous observations of non-linearities with semi-conductors as shown in figure 3.56 in \cite{peinke2012encounter}.

\section{Summary and conclusion}

In this study, we investigated the impact of side-to-side and fore-aft harmonic motions on the wake of a model floating wind turbine in laminar wind. Our research focused on the impact of motion frequency on the wake, with a constant amplitude of movements of $\sim 0.01D$. This paper presents a number of new findings. First, our results demonstrate the equivalence between surge and pitch, as well as, between sway and roll (see figures \ref{fig:sway_roll} and \ref{fig:surge_pitch} in \S \S \ref{sub-section:equivalence_DOF}). Therefore, we focused our research on sway DoF for side-to-side motions and surge DoF for fore-aft. Second, the experimental results are in excellent agreement with the CFD simulations from \cite{li2022onset} about the wake of a swaying turbine in laminar wind (figure \ref{fig:recov_sway_st}). This implies that the wake of the floating turbine does not depend on the Reynolds number, at least for $\Rey > 10^5$. Third, our findings indicate that both types of movement significantly enhance wake recovery, especially for $St \in [0.3, 0.6]$ with sway (figure \ref{fig:recov_sway_st}) and $St \in [0.3, 0.9]$ for surge (figure \ref{fig:recov_surge_st}). Fourth, we worked out different wake modes depending on the direction of motions and $St$. For low $St$ and surge DoF we found a new mode, namely a pulsing mode of the wake. For the same DoF and higher $St$, a self generated meandering mode is observed. These are all non-linear dynamical effects.
\\  
\\
 At such low inflow turbulence, the perturbations induced by the movements of the floating wind turbine are responsible of different dynamics of the wake, which are particularity  visible in the shear region of the wake ($y \approx R$). On the one hand, sway motions are  inducing large sideways wake meandering, especially for $St \in [0.2, 0.5]$. The wake synchronises (pseudo lock-in) to the frequency of movements as seen in the spectrum of the wind fluctuations, which shows a clear and narrow peak at $f_p$ (see figure \ref{fig:psd_sway} (e, f)). The small perturbations induced by the movements are largely amplified and results into clear meandering at $f_p$ (see figure \ref{fig:sway_meandering} (c,f)). This dynamical response is a typical non-linear behaviour. Similarly, for $St \in [0.25, 0.55]$, the wake of the surging turbine shows pulsating movements at the excitation frequency (see figure \ref{fig:surge_meandering} (b)). For both DoFs, the broadband frequency range of the wake of the fixed turbine (figures \ref{fig:psd_sway} (d) and \ref{fig:psd_surge} (d)) is slaved to a single frequency, driven by the motions. This monochromatic response is in some ways much easier to describe than the more complex broadband spectra of the fixed turbine which contains a wide range of flow structures.
\\
On the other hand, for $St > 0.55$, the wake of the surging turbine is experiencing meandering at a self generated frequency, $f^*$, in the range of the natural frequency of the wake of the fixed turbine ($f^* \in [0.1, 0.5]U_{\infty} / D$). For these cases, the wake's dynamics are more complex, the spectrum shows peaks at $f_p$ and $f^*$ as well as mixing components (figure \ref{fig:psd_non_linear} (g, h)). This is an explicit proof of non-linear interactions of these two frequencies, leading to so-called quasi periodicity. In summary, we worked out one dynamic mode for sway (meandering) and two dynamic modes for surge (pulsating and meandering).
\\
\\
These dynamic behaviours play an important role in the recovery process and provide insights into the results observed for the average values. Most interestingly, we see that the synchronisation of the broadband wake modes to the driving frequency leads to a faster recovery. This result seem on a first view counter-intuitive. On the one hand, synchronisation makes the wake structures more organised (i.e. more coherent). In similarity with the shielding effect of tip vortices demonstrated by \cite{lignarolo2015tip}, we could expect less recovery. On the other hand, the pulsating and meandering structures in the wake of the floating wind turbine do not prevent momentum exchange with the flow as do tip vortices, but rather enhance momentum transport and increase the exchange area, thus improving recovery from the wake centre. In a similar way, it is astonishing that the more complex but still low dimensional quasi-periodic states for the surge case lead to an even greater enhancement in recovery (the recovery is about 10 \% more for $St \approx 0.8$ compared to $St \approx 0.4$). Both sideways and fore-aft motions enable the far-wake to be developed closer to the rotor, at $x \leq 6D$ for the most impactful cases, while it is observed at $x \approx 10D$ for the fixed case, which is in close relation to the dynamic behaviour of the wake. This study shows that non-linear dynamic phenomena must be taken into account to better understand the behaviour of the wake, especially in the transition region (here $x \in [3D, 6D]$).
\\
\\
The results of the paper demonstrate that perturbations in the near-wake due to platform motions are advected downstream, amplified and interact non-linearly which results in large meandering or pulsating movements in the wake. Interestingly, these motions could enhance the performance of a wind farm by increasing the energy available to downstream turbines. However, it is important to note that the large coherent structures could also contribute to increased loadings on downstream turbines. It is an open question whether or not these features will reduce turbine spacing within future floating wind farms.
\\
\\
\textbf{Acknowledgement:} This work has received funding from the EU’s Horizon 2020 research and innovation programme under the Marie Sklodowska-Curie grant agreement N$^{\circ}$860879 as part of the FLOAWER
consortium. The authors would like to thank Julian Jüchter, Jannis Maus, Dr. Jaroslaw Puczylowski and Agnieszka Hölling for their valuable help with the experiments.
\\
\\
\textbf{Declaration of Interests:} The authors report no conflict of interest.

\newpage

\appendix
\section{Additional data to the experiments}
\label{appA}

\begin{table}
  \begin{center}
\def~{\hphantom{0}}
  \begin{tabular}{lccc}
      Parameter  & Notation   &   Value & Unit \\[3pt]
       Rotor diameter  & $D$ & 0.58 &  $m$\\
       Rotor radius  & $R$ & 0.29 &  $m$\\
       Hub height   & $H_{hub}$ & 0.96 &  $m$\\
       Tower length  & $T_{length}$ & 0.45 &  $m$\\
       Cut-in wind speed  & $U_{in}$ & 2.5 &  $~m.s^{-1}$\\
       Rated wind speed   & $U_{rated}$ & 7.5 & $~m.s^{-1}$\\
       Rated power & $P_{rated}$ & 25.4 & W\\
       Rotational speed & $\omega$ & 400-1600 & rpm\\
       Tip speed ration & $TSR$ & 4-9 & [-]\\
  \end{tabular}
  \caption{MoWiTO 0.6 characteristics}
  \label{tab:mowito_06_table}
  \end{center}
\end{table}

\begin{table}
  \begin{center}
\def~{\hphantom{0}}
  \begin{tabular}{lcccccccc}
  Test  & DoF   &   Amplitude & $A^*$& Frequency & Wind Speed & $St$ & $C_T$ & $\Rey$\\[3pt]
    [-]  & [-]    &   [$mm$] or [$^{\circ}$] & [-] & [$H_z$] & [$m/s$] & [-] & [-] & $\times 10^5$\\[3pt]
    \hline
   A.1  & Fixed   &  0 & 0 &  0 & 5 & 0 & 0.81 & 2.3\\
  A.2  & Sway   &  4 mm & 0.007 &  3.3 & 5 & 0.39 & 0.81 & 2.3 \\  
  A.3  & Roll   &  0.5$^{\circ}$ & 0.007 &  3.3 & 5 & 0.39 & 0.81 & 2.3 \\  
  A.4  & Sway   &  38 mm & 0.065 &  0.3 & 5 & 0.03 & 0.81 & 2.3 \\  
  A.5  & Roll   &  5$^{\circ}$ & 0.065 &  0.3 & 5 & 0.03 & 0.81 & 2.3 \\ 
     \hline 
     B.1  & Fixed   &  0 & 0 &  0 & 5 & 0 & 0.81 & 2.3\\
  B.2  & Surge   &  4 mm & 0.007 &  3.3 & 5 & 0.39 & 0.81 & 2.3 \\  
  B.3  & Pitch   &  0.5$^{\circ}$ & 0.007 &  3.3 & 5 & 0.39 & 0.81 & 2.3 \\  
  B.4  & Surge   &  38 mm & 0.065 &  0.3 & 5 & 0.03 & 0.81 & 2.3 \\  
  B.5  & Pitch   &  5$^{\circ}$ & 0.065 &  0.3 & 5 & 0.03 & 0.81 & 2.3 \\  
     \hline
   C.1  & Fixed   &  0 & 0 &  0 & 5 & 0 & 0.81 & 2.3\\
   C.2  & Sway   &   4 mm & 0.007 & 1-5 & 5 & 0.12-0.58 & 0.81 & $2.3$\\
    \hline
    D.1  & Fixed   &  0 & 0 &  0 & 3 & 0 & 0.76 & 1.4\\
   D.2  & Surge   &  4 mm & 0.007 &  1-5 & 3 & 0.19-0.97 & 0.76 & 1.4\\
   D.3  & Fixed   &  0 & 0 &  0 & 5 & 0 & 0.81 & 2.3\\
   D.4  & Surge   &  4 mm & 0.007 &  1-5 & 5 & 0.12-0.58 & 0.81 & 2.3\\
  \end{tabular}
  \caption{Cases investigated. \textit{For all cases, $TSR \approx 6.0$. \\
For rotational DoFs: $A^* = T_{length} \times tan(A)$.}}
  \label{tab:cases_investigated}
  \end{center}
\end{table}

\begin{figure}
  \centerline{\includegraphics[width=15cm]{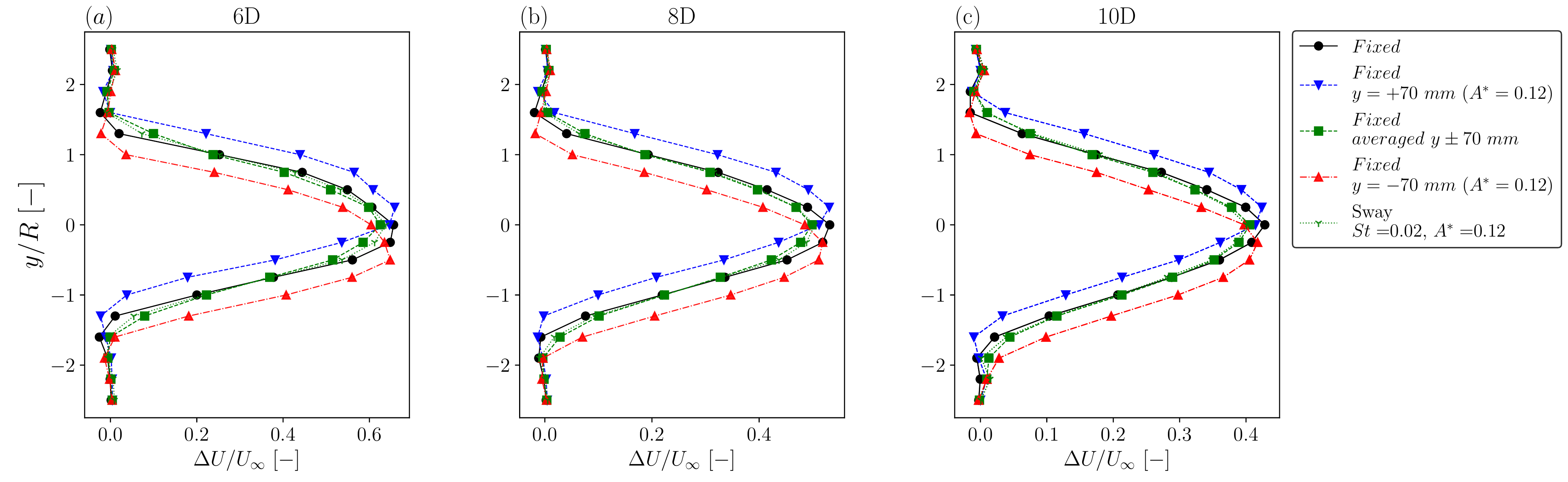}}
  \caption{Wake deficit profiles at 6D (a), 8D (b), 10D (c) for: fixed case, fixed with an offset of $\pm 0.12D $ (70 mm) in $y$ (horizontal), average of the offset cases and sway with $A^* = 0.12,~St = 0.02$, $Re = 2.3 \times 10^5$}
\label{fig:sway_superposition}
\end{figure}

\newpage

\bibliographystyle{jfm}
\bibliography{jfm_TM_wake_fowt_1_0}

\end{document}